\newif\ifarxiv
\arxivtrue 

\ifarxiv
	\documentclass[10pt]{article}
	\usepackage[margin=1in]{geometry}
	\usepackage[slantedGreek]{mathpazo}
	\usepackage[pdftex]{graphicx}
	\usepackage{amsfonts}
	\usepackage{amssymb}
	\usepackage{amsthm}
	\usepackage{graphicx,float}
	\usepackage{amsmath}%
	\usepackage{hyperref}
	\usepackage{natbib}
	\usepackage{bm}
\else
\fi

\usepackage{color,soul,marginnote}

\DeclareMathOperator*{\maxf}{maximize \quad}
\newcommand{\maxdisp}[3]{\[\maxf_{#1} #2\quad \mbox{subject to}\quad #3\]}

\newcommand{\cov}{\mathrm{cov}}

\ifarxiv
\title{Merging Two Cultures: Deep and Statistical Learning}
\author{
	Anindya Bhadra\thanks{Purdue University, bhadra@purdue.edu}, \; Jyotishka Datta\thanks{Virginia Tech, jyotishka@vt.edu}, \; Nick Polson\thanks{U Chicago. ngp@chicagobooth.edu}, \; Vadim Sokolov\thanks{George Mason. vsokolov@gmu.edu} \; and \; Jianeng Xu\thanks{U Chicago. jianeng.xu@chicagobooth.edu}}
\date{First Draft: Nov 1, 2019\\
This Draft: Oct 12, 2021
}
\else
\fi

\graphicspath{{../fig/}}

\begin{document}
\ifarxiv
\maketitle
\begin{abstract}
\noindent
Merging the two cultures of deep and statistical learning provides insights into structured high-dimensional data. Traditional statistical modeling is still a dominant strategy for structured tabular data. Deep learning can be viewed through the lens of generalized linear models (GLMs) with composite link functions. Sufficient dimensionality reduction (SDR) and sparsity performs nonlinear feature engineering. We show that prediction, interpolation and uncertainty quantification can be achieved using probabilistic methods at the output layer of the model. Thus a general framework for machine learning arises that first generates nonlinear features (a.k.a factors) via sparse regularization and stochastic gradient optimisation and second uses a stochastic output layer for predictive uncertainty. Rather than using shallow additive architectures as in many statistical models, deep learning uses layers of semi affine input transformations to provide a predictive rule. Applying these layers of transformations leads to a set of attributes (a.k.a features) to which predictive statistical methods can be applied. Thus we achieve the best of both worlds: scalability and fast predictive rule construction together with uncertainty quantification.  
Sparse regularisation with un-supervised or supervised learning finds the features. We clarify the duality between shallow and wide models such as PCA, PPR, RRR and deep but skinny  architectures such as autoencoders, MLPs, CNN, and LSTM. The connection with data transformations is of practical importance for finding good network architectures.  By incorporating probabilistic components at the output level we allow for predictive uncertainty. For interpolation we use deep Gaussian process and ReLU trees for classification. We provide applications to regression, classification and interpolation. Finally, we conclude with directions for future research. 

\vspace{0.1in} 

\end{abstract}
\else
\fi



\section{Introduction}	
Model specification is one of the most challenging parts of statistical modeling as originally  discussed in Fisher's seminal paper \citep{fisher1922mathematical}.
Breiman \citep{breiman2001statistical} highlighted the contrast between an algorithmic  approach and traditional statistical modeling for 21st century data analytics.  Algorithmic approaches focus their effort on understanding high-dimensional data structure. Deep learning is an  algorithmic modeling approach which has changed the landscape for text \citep{devlin2018bert} and image analysis \citep{litjens2017survey} and many other areas of applications  \citep{bhadra2019horseshoe,heaton2017deep,dixon2017}. Our goal is to show that deep learning has  wide applicability to traditional statistical areas for tabular data structures including categorical, spatial and time series analysis.  Until now, traditional statistical models have relied heavily on additive functions with low approximation capacity based on shallow architectures. From a statistical viewpoint, much attention has been paid to stochastic models that combine with the deterministic part of a statistical model. 

Deep Learners are based on superposition of univariate affine functions 
\citep{polson_deep_2017} and are universal approximators. Whilst Gaussian Process \citep{gramacy2008bayesian,higdon2008computer} are also universal approximators and can capture relations of high complexity, they typically fail to work in high dimensional settings. Tree methods can be very effective in high dimensional problems. Hierarchical  models are flexible stochastic models but require high-dimensional integration and MCMC simulation. Deep learning, on the other hand, is based on scalable fast gradient learning algorithms such as Stochastic Gradient Descent (SGD) and its variants.  Modern computational techniques such as automated differentiation (AD) and accelerated linear algebra (XLA) are available to perform stochastic gradient descent (SGD) at scale within  TensorFlow or PyTorch, thus avoids the curse of dimensionality by simply pattern matching and using interpolation to predict in other regions of the input space. The algorithmic culture has achieved much success in high dimensional problems. DL assumes a very flexible class of predictors, $f(x)$, and directly train this predictor using a predictive mean squared error loss. Such classes of functions include decision trees and neural networks. The goal is simple to find a predictor rule. Can we find a good predictor (a.k.a. algorithm) $f(x)$ to evaluate on $x$ to predict output $y$? The caveat with an algorithmic approach is that it lacks uncertainty quantification. 

The statistical modeling approach makes uncertainty quantification paramount and, following Breiman, we write 
$$            
\mathrm{output} = f(\mathrm{predictor \; variables}, \mathrm{random \; error} , \mathrm{parameters})
$$
The limitations of statistical modeling are clear as model specification and validation is hard particularly in high dimensions.

\paragraph[]{Interpolation}
Gaussian Process and piece-wise polynomial functions \citep{wahba1990spline} are popular approaches to interpolate and require specifying smoothness parameters or learning those using MLE or Bayesian inference. For example, often Gaussian Process models are used to quantify uncertainty of complex scientific simulators \cite{higdon2008computer, gramacy2008bayesian} or complex geo-spacial processes \cite{kim2005analyzing}. We show how they can be merged with deep learning.

\cite{neal2012bayesian} has shown that a function space defined by Bayesian neural network with Gaussian weights approximates is a Gaussian process as the number of neurons goes to infinity. Later \cite{mackay1998introduction} has argued that Gaussian process models should be preferred since they do not require specifying architecture and only priors for hyperparameters of correlation functions need to be defined.  On the other hand, modeling non-stationary and non-isotropic data is hard to model with Gaussian process, while neural networks can handle those types of relations. Thus, the fact that an architecture with specific distribution over weights approaches Gaussian process does not necessarily mean that GP is to be preferred. Several approaches were proposed to address the problem of modeling non-stationary and non-isotropic data with GP. \cite{gramacy2008bayesian} proposes using decision trees to partition the data so that each partition constrains a stationary subset and then a separate GP model is used for each subset. \cite{fadikar2018calibrating} use several Gaussian Process models for different quartiles of the time series data to model non-stationary epidemic data. \cite{srivastava2014dropout} is a regularization technique that sets weights to randomly to zero. 

Our approach then adds to traditional deep learning by incorporating probabilistic components at the output level, given learned data filters, which then allow for predictive uncertainty.. We illustrate the merging of the two cultures  using deep Gaussian process and high dimensional classification using ReLU trees which provide an alternative to  random forests. Finally, we outline directions for future research.

\paragraph{Uncertainty Quantification.} Statistical models are capable of representing uncertainty in predictions and parameters. When input-output relations are modeled using deterministic functions, such as ridge or neural network, this property is lost. Understanding uncertainty is key in many science and engineering applications. 
Neural networks are essentially a nonparametric regression which uses sparse or skinny representation \citep{olshausen1996emergence}.
\cite{bishop1995neural} considers neural networks from statistical modeling p	oint of view and presents neural networks as an extension to more traditional functions used in statistical modeling. 

\cite{ripley2007pattern} presents feed forward neural network as a way to project input into lower dimensional space within which the approximation can be preferred and compares is to projection pursuit regression. \cite{bengio2009learning} also presents a neural network as a way to generate representation of the data in a different subspace. Each layer is a representation. 

The rest of the paper is outlined as follows. Section \ref{sec:merging}  merges the two cultures of statistical modeling and deep learning and describes classes of deep learners. To do this, we first describe how to perform unsupervised and supervised dimension reduction of the input space. Hand-crafted as well as learned transformations are discussed. Our approach differs from traditional deep learning by using probabilistic model as the last output layer. This allows us to use traditional statistical uncertainty quantification methods such as in logistic regression \citep{polson2013bayesian} or Gaussian Process \citep{gramacy2008bayesian}. Section \ref{sec:predictors} discusses the problem of model selection.  Section \ref{sec:applications} provides applications in regression, classification and interpolation. We provide a deep learning alternative to treed models. Finally, Section \ref{sec:discussion} concludes with directions for future research.
 
\subsection{General latent feature model}
Given a training dataset of input-output pairs $ ( Y_i , X_i )_{i=1}^N $ the goal is to find a prediction rule for a new output $ Y_* $ given a new input $ X_*$.  Let $Z$ denote latent hidden features that are to be hand-coded or learned from the data and our nonlinear latent feature predictive model takes the form 
\begin{align}
Y\mid Z & \sim p(  Y \mid Z ) \label{eq:bry}\\ 
Z &=\phi(X) \label{eq:brf}
\end{align}
where $\phi$ is a data transformation, that is allows for relations between latent features $Z = \phi(X)$ and $Y$ to be modeled by a well understood probabilistic model $p$. Typically $\phi$ will perform dimension reduction or dimension expansion and can be learned from data. The top level of our model is stochastic. This gives a full representation of the predictive uncertainty in predicting new $ Y_\star $. In many cases, our predictor is simply the conditional mean $ \hat Y = E(Y\mid F ) $.


\paragraph{Finding Predictions.} The key is to find a good data tranformation, e.g. deep learning architecture, that predicts well. Essentially DL solves two problems. First it finds the latent features $Z$. Second, it interpolates/predict a new output for a new input $X_*$. This latter property is governed by predictive cross-validation. Predictive uncertainty is also required, which is provided by our probabilistic model at the top layer $p(Y\mid Z)$.




\section{Merging Deep and Statistical Learning}\label{sec:merging}
The main role of feature selection is to find the data transformation so that the relations between input and output can be captured by one of the statistical models. For example we often apply log-transformations so that a linear model can be used. In this section we discuss several approaches to data transformation that enable usage of simple predictive rules for the transformed data.  There are two forms of data transformations, first dimension expansion and second dimension reduction.

\subsection{Dimensionality Expansion}
First, we review dimensionality expansion data transformation that transforms input vector $x$ into higher dimensional vector $\phi(x)$. One approach is to use hand-coded predictors. This expanded set can include terms such as interactions, dummy variables or nonlinear functional of the original predictors.  The goal is to model the joint distribution of outputs and inputs, namely $ p( y , \phi(x)) $, where we allow or stochastic predictors.

The joint distribution $ p( y , \phi(x)) $ is characterized by its two conditionals distributions 
\begin{itemize}
	\item  $ p( y | \phi(x) )  $.  This performs prediction via $ y = f(x) = E(y|x) $ and uncertainty quantification using its probabilistic structure.
	\item  $ p(  \phi(x) | y ) $.  This conditional is very dimensional and we need to perform dimension reduction and find an efficient set of nonlinear features to be used as predictors in step 1. Selection using sparsity and deep learners will be the methods used.
\end{itemize}

\begin{figure}[H]
	\centering
	\includegraphics[width=0.7\linewidth]{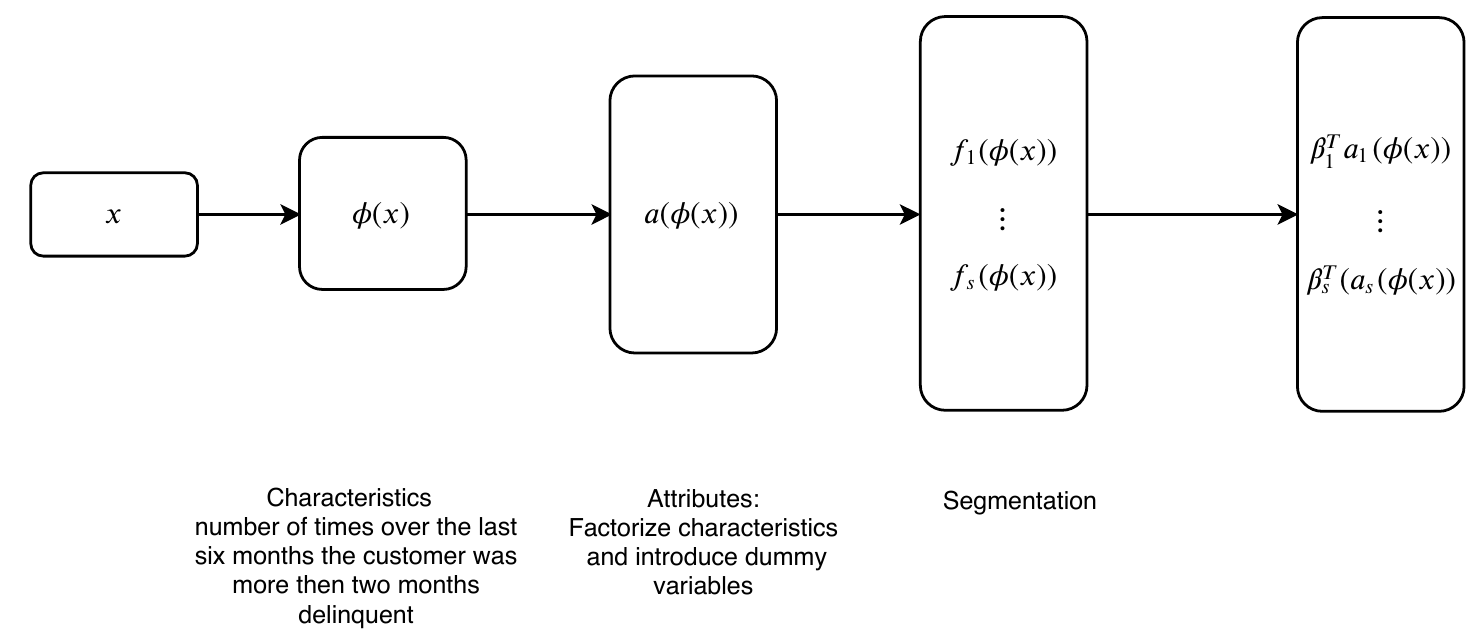}
	\caption{Segmented Score Card Logistic Regression Architecture used by Fair, Isaac to predict delinquency}
	\label{fig:fico}
\end{figure}

As an example of this approach, we describe a predictive model used by Fair, Isaac to predict credit worthiness \citep{Hoadley2000}. Today this indicator is known as FICO score. The overall architecture of the FICO model is shown in Figure \ref{fig:fico}.

In the Fair, Isaac architecture, first the $x$ variable (monthly bills and payments over the last 12 months) was transformed into many interpretable variables, e.g. months delinquent and then calculate the \textbf{characteristics} vector $\phi(x)$ from those  interpretable variables. For example, the number of times in the last six months that the customer was more than two months delinquent. Than the data was segmented and a separate predictive model developed for each of the segments. Each segment-specific predictive model was a variant of the logistic regression.

\paragraph{Kernel Expansion} The idea is to enlarge the feature space via basis expansion. The basis is expanded using nonlinear transformations of the original inputs $\phi(x) = (\phi_1(x),\phi_2(x),\ldots,\phi_M(x))$ so that linear regression $\hat y = \phi(x)^T\beta + \beta_0$ or generalized linear model can be used to model the input-output relations. Kernel trick increase dimensionality, and allows hyperplane separation. The transformation $\phi(x)$ is specified via a kernel function $K$ which calculates the dot product of feature mappings
\[
K(x,x') = \phi(x)^T\phi(x').
\]
By choosing feature map $\phi$, we implicitly choose a kernel function. 
For example, when $x\in \mathbb{R}^2$, choosing $K(x,x') = (1+x^Tx')^2$ is equivalent to expanding the basis to $\phi(x) = (1,\sqrt{2}x_1, \sqrt{2}x_1, x_1^2,x_2^2,\sqrt{2}x_xx_2)$.

\begin{figure}[H]
\centering
\includegraphics[width=\linewidth]{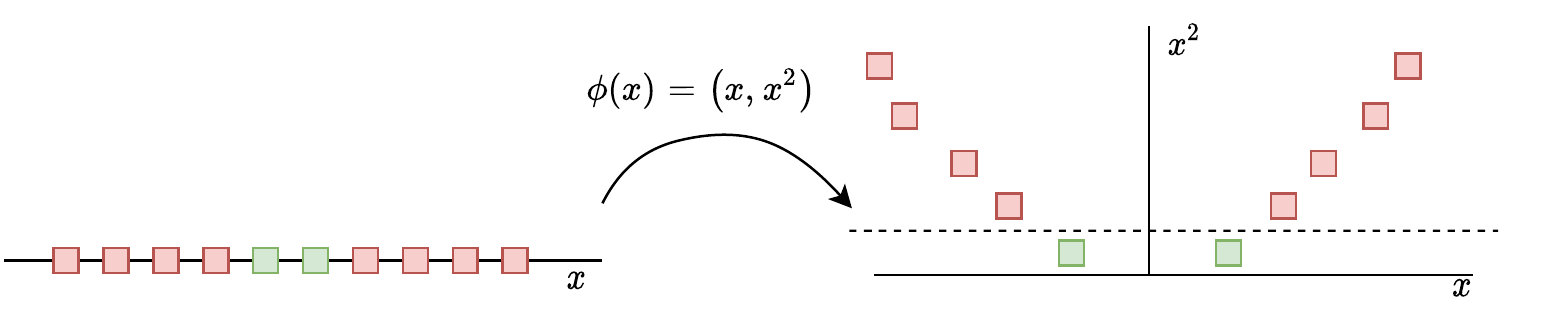}
\caption{}
\label{fig:}
\end{figure}

\paragraph{Tree Expansion} Similar to kernels we can think of trees as a technique for expanding a feature space. Each region in the input space defined by a terminating node of a tree correspond to a new feature. Then predictive rule becomes very simple, identify in which region the new input is and use average across observations from this region to calculate the prediction.

\paragraph{Deep Learning Expansions}
Similar to a tree model that finds features (a.k.a tree leaves) by splitting the input space into rectangular regions, the deep learning model finds the regions by using hyperplanes at the first layer and combinations of hyperplanes in the further layers. The prediction rule is embedded into a parameterized deep learner, a composite of univariate semi-affine functions, denoted by $F_W$ where $W = [ w^{(1)}, \ldots , w^{(L)} ] $ represents the weights of each layer of the network.  A deep learner takes the form of a composition of link functions
$$
F_W = f_1 \circ \dots \circ f_L  \; {\rm where} \; f_L = \sigma_L ( w_L \phi (x) + b_L) 
$$
where $ \sigma_L $ is a univariate link function. Specifically, let $z^{(l)} $ denote the $l$-th layer, and so $x = z^{(0)}$.
The final output is the response $y$, which can be numeric or categorical.
A deep prediction rule is then
\begin{align*}
z^{(1)} & = f^{(1)} \left ( w^{(0)} \phi(x) + b^{(0)} \right ),\\
z^{(2)} & = f^{(2)} \left ( w^{(1)} z^{(1)} + b^{(1)} \right ),\\
& \ldots\\
z^{(L)} & = f^{(L)} \left ( w^{(L-1)} z^{(L-1)} + b^{(L-1)} \right ),\\
\hat{y} (x) & = w^{(L)} z^{(L)} + b^{(L)}\,.
\end{align*}

\begin{figure}[H]
	\begin{tabular}{cp{0.5\linewidth}}
	\includegraphics[width=0.5\linewidth]{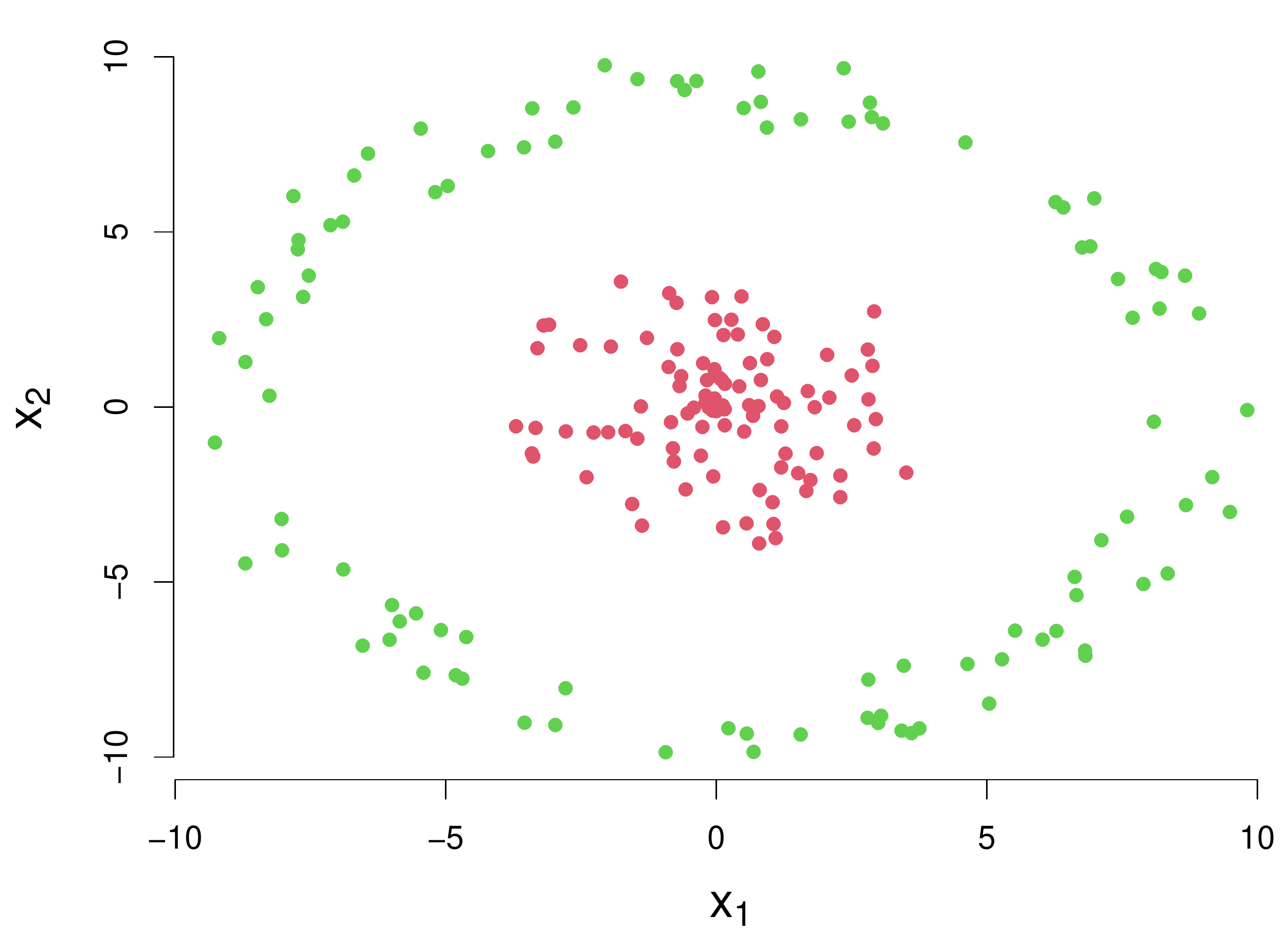} & \includegraphics[width=\linewidth]{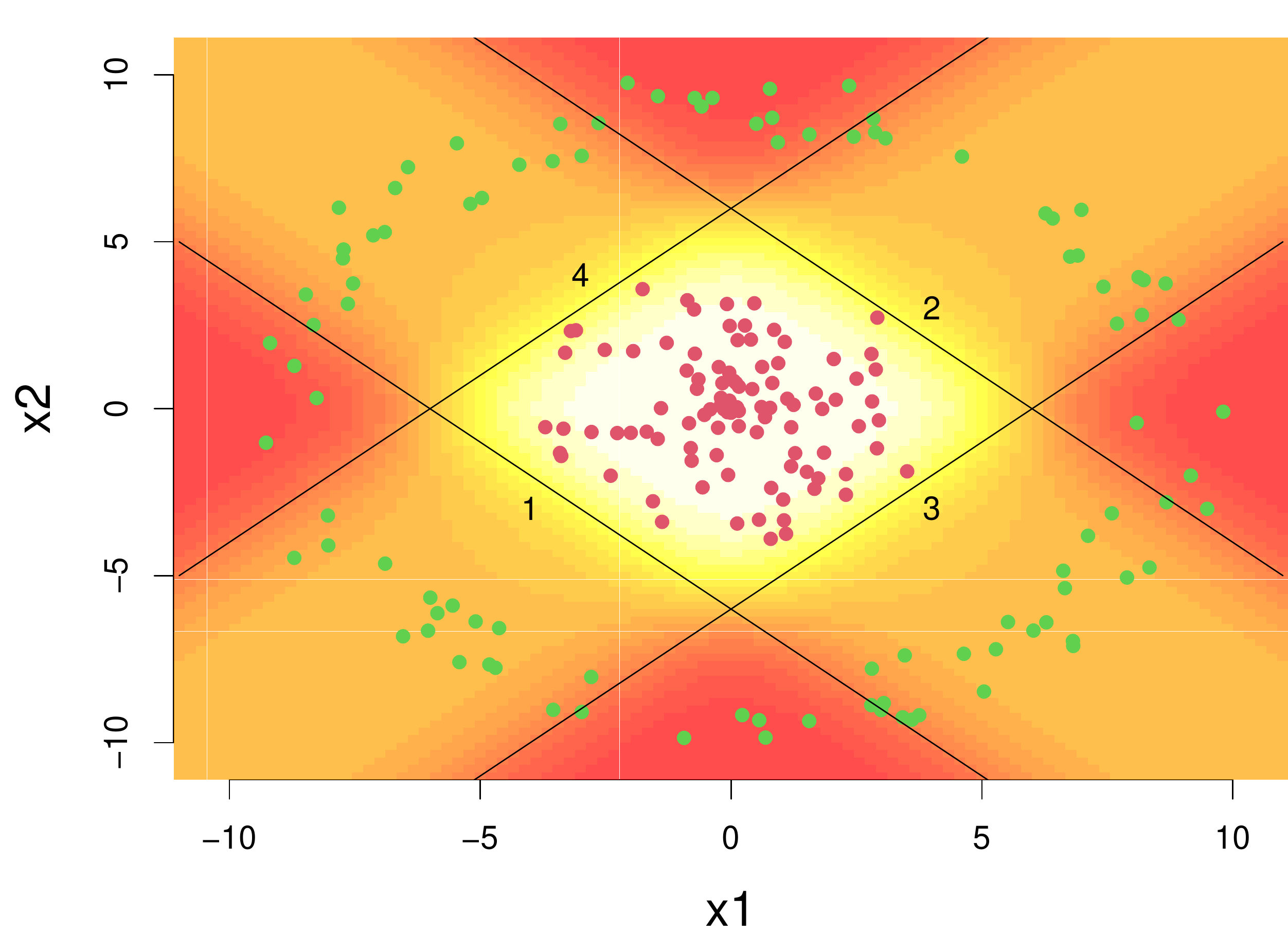}\\
	(a) Dataset & (b) Hyperplanes and probability heat map for classification based on the four hyper-planes
\end{tabular}
	\caption{Training data set and four hyper-planes defined by hidden layer of a neural network}
	\label{fig:nn-circle2}
\end{figure}

We demonstrate the similarity between a tree model and a layer of a deep learning model by showing how a DL model classifies observations form a simulated daugnat dataset shown in Figure \ref{fig:nn-circle2}(a). It is clear that a linear separator will no work in this case, however, we can use four lines to split the data into nine regions as shown in Figure \ref{fig:nn-circle2}(b). 

Given this split, it is easy to design a predictive rule for classifying the red and green dots. If a point is on the left of lines 2, 4 and on the right of lines 4, 1, then classify as red and classify as green otherwise. The four lines are given by the linear system 

\[
a = \left(\begin{array}{rr}
	1 & 1\\
	-1 & -1\\
	-1 & 1\\
	1 & -1
\end{array}\right)\left(\begin{array}{rr} x_1 \\ x_2
	
\end{array}\right), \qquad z = \sigma(a)
\]
Finally, we perform classification by a logistic regression
\begin{align*}
	&\mu =  -3 + z_1 + z_2 + z_3 + z_4\\
	&P(y=1\mid x) = e^{\mu}/(1+e^{\mu}).
\end{align*}

Figure \ref{fig:nn-circle2}(b) shows the heat plot of the $P(y=1\mid x) $ with red being 1 and white being 0. 


We can also visualize the deep learning model using a tree-like diagram to highlight the similarity between two approaches. 
\begin{figure}[H]
	\includegraphics[width=1\linewidth]{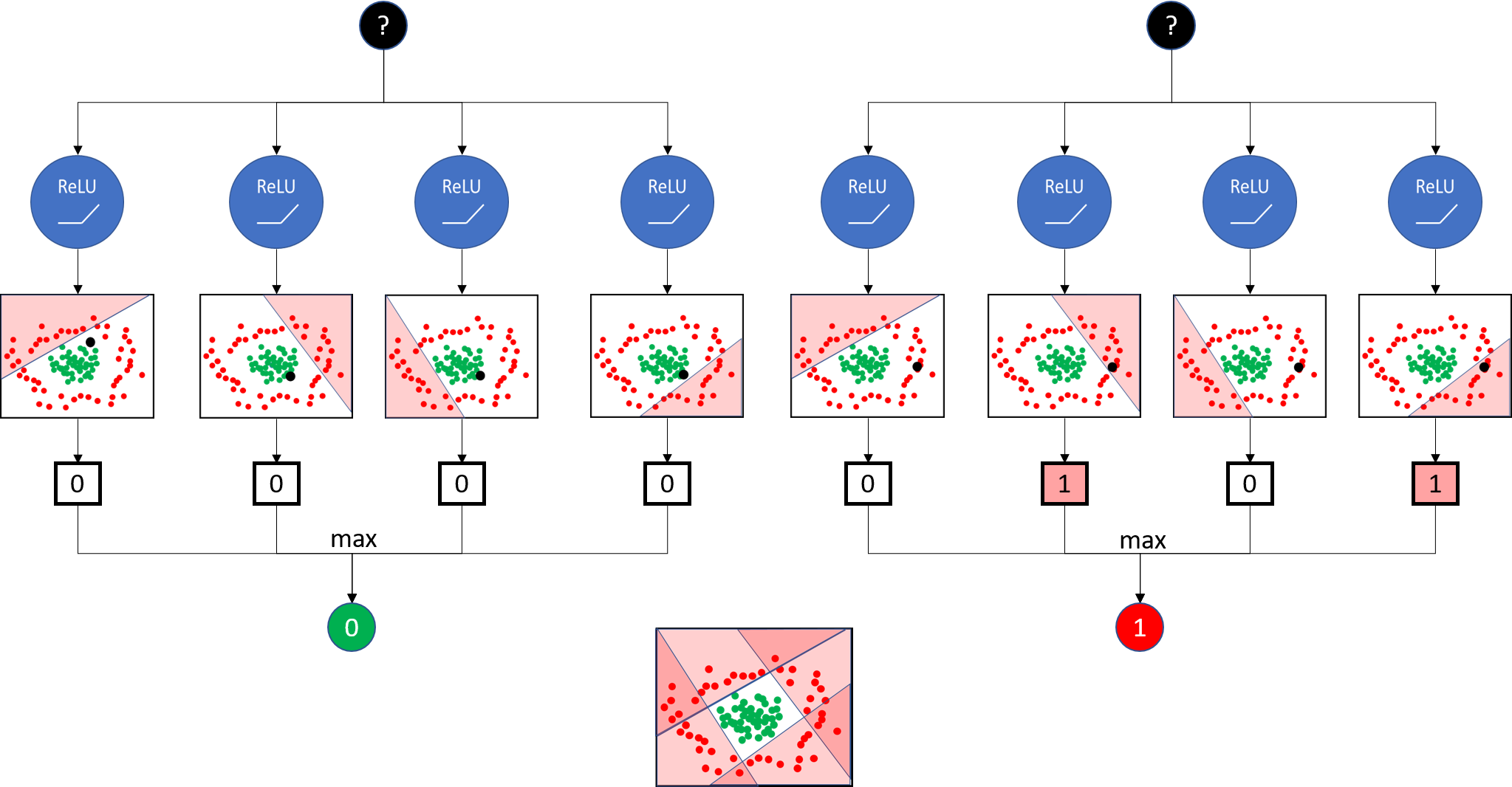}
	\caption{Deep ReLU network}
	\label{fig:dl-tree}
\end{figure}


It is often beneficial to replace the original input $X$ with the features $Z = g(X)$ of lower dimensionality when developing a predictive model for $Y$. It was shown, for example, in the context of regressions, that a lower variance prediction rule can be obtained in lowe dimensional space \citep{kofi09} . For fully Bayesian discussion see \cite{lindley68}.

To achieve good generalisability we need to be able to perform nonlinear dimension reduction and to find a suitable set of features/factors. Deep learners together with sparse optimization provides such a framework. 
From a probabilistic view point, it is natural to view input-output paris as being generated from some distribution
\[
(y_i,x_i) \sim p(y,x), \; i=1,\dots,N.
\]
We are interested in two conditional distributions
\begin{itemize}
	\item $p(y\mid x)$ the probabilistic quantification of uncertainty about the output $y$ at a new input $x$. One of the key assumptions--- an equivalent to sufficient data reduction in a statistical context-- is to assume that 
	$$
	p(y\mid x) \text{ is equivalent to  } p(y \mid F_W\left(\phi(x))\right).
	$$
	Here $\phi(x)$ is a dimension increasing set of characteristics set, e.g. the inclusion of dummy variables (one hot encodings), interaction terms, and $F_W$ is dimension reducing deep learning model, trained from input-output pairs. 
	
	Typically $ F_W(\phi(x))$ is constructed as a combination of hand-crafted and learned transformation from deep learning.  This provides the efficient data reduction necessary to provide high dimensional prediction. 
	\item $p(x\mid y)$ the conditional distribution of inputs, given the outputs which can be used to perform supervised learning of data transformations comprising the deep learner $F_W$. In the unsupervised case, one simply used the marginal distribution $p(x)$.
\end{itemize}
Figure XXX shows how the data transformations work to construct the nonlinear deep learners.
DL simply uses a composition/superposition of semi-affine filters (a.k.a. link functions).

This leads to the following framework for high-dimensional modeling. 

Given a training data $\{(y_i,x_i)\}_{i=1}^n$, $x_i \in \mathbb{R}^p$ we first use
$ p( x| y ) $ to uncover structure in the predictors relevant for modeling the output $y$.
The learned factors are denoted by $ F(\phi(x))$ and are constructed as a sequence of input filters. Finally, the predictive model is given by a probabilistic model of the form 
$ p(y|x) \equiv p( y | F(\phi(x)))$.
Here $\phi: \mathbb{R}^p \rightarrow \mathbb{R}^c,~c \gg p$ initially expands the dimension of the input space by including terms such as 
interactions, dummy variables (a.k.a. one hot encodings) and other nonlinear features of the input space deemed relevant.  Then $F$ reduce dimension by deep learning by projecting back with univariate activation function (a.k.a. link) into an affine space (a.k.a regression)

Deep learning can then be viewed as  a feature engineering solution and one of finding 
nonlinear factors via supervised dimension reduction.
A composition of hand-coded characteristics--dimension expanding--- with
supervised learning of data filters--dimension reduction
Advances in computation allow for massive data and gradients of high dimensional nonlinear filters Neural networks can be viewed from two perspectives. Either as a flexible link function in a GLM model \cite{mccullagh2019generalized} or a method to achieve dimensionality reduction, similar to sliced inverse regression \cite{li1991sliced} or sufficient dimensionality reduction \cite{cook2009likelihood}.

This framework also sheds light on how to build deep (skinny) architectures. Given $n$ data points, we split into $L = 2^p$ regions \citep{harding1967number} so that there is a "fixed" sample size within each bin.  To summarize 

\begin{itemize}
	\item Transform $x$ into many interpretable characteristics. First generate several time series from raw inputs, e.g. months delinquent. Then extract futures from those time series, e.g. number of times over the last six months the customer was more then two months delinquent. This process leads to thousands of characteristics that are screened to be included into the predictive models
	\item Then population is segmented based on the screened characteristics. This process was manual. Similar to the CART algorithm. 
	\item A separate function $f(x)$ (scorecard) was developed for each segment. Each characteristic was binned into sets called attributes. A scorecard is linear function of the attribute indicators (dummy) variables. 
\end{itemize}

Increase of dimensionality is the key! 24 inputs led to thousand of characteristics and hundreds after screening. After discrediting input about 10 attributes per characteristics and after introducing 10 segments, we get tens of thousands of features. Essentially it is a generalized additive models (GAM) with bin smoothing. 

One advantage of ``depth'' is that the hierarchical mixture allows the width of a given layer to be manageable. With a single layer (e.g., kernel PCA/SVM) we need exponentially many more basis functions in that layer. Consider kernel PCA with say RBF kernels: technically there are infinitely many basis functions, but it cannot handle that many input dimensions.  Presumably a deep neural network allows a richer class of covariances that allows anisotropy, nonstationarity etc. In the end, this is reflected in the function realizations from a DNN. To see this, consider the deep GP models, which are infinite width limits of DNNs (ref). There is a recursive formula connecting the covariance of layer $k$ to that of layer $k+1$, but no closed form. The covariance function of the final hidden layer is probably very complicated and capable of expressing a lot of features, even if the covariances in each layer may be simple. The increase in dimensionality happens through the hierarchical mixture rather than trying to do it all in one layer.

From a statistical viewpoint, this is similar to the liner shallow wide projections introduced  Wold (1955) and the sufficient dimension reduction framework of  Cook (2007).

In an unsupervised learning context we simply use information in the marginal distribution, $p(x)$,  of the input space as opposed to the conditional distribution, $p(x|y)$.
PCA, PCR, RRR, PPR all fall into this category. PLS, SIR are examples of supervised learning of features, See \citep{polson_deep_2017} for further discussion.

The joint distribution $ p(y,x) $ is characterized by its two conditionals $  p( y |  x ) $ and $ p( x | y ) $.  We first embed $x$ into $ \phi(x) $ which can be very high dimensional--- including dummies, interactions etc.  Then we need to model $p( y , \phi(x))$ and this requires dimension reduction.  We will perform this using a deep learner $F_W$ where $W$ indexes the weights in each affine layer. 

Estimate $\hat{W}$ and hence find the nonlinear feature extraction $F_{\hat{W}}$  using  
stochastic gradient descent. 
inverse conditional distribution
$ p( x | y ) $. Data reduction techniques

\subsection{PCA, PCR and SVD Algorithm}
Given inputs $x$ and outputs $y$ and associated observed data $X\in R^{n \times p}$ and $Y \in R^{n\times q}$. The goal is to find data transformations $(Y,X) = \phi(Y,X)$ so that modeling the transformed data becomes an easier task. In this paper we consider several types of transformations and model non-linear relations. 

We start by reviewing widely used singular value decomposition (SVD) which allows to find linear transformations to identify lower dimensional representation of $X$, the technique is known as principle component analysis (PCA) or both $X$ and $Y$, known as partial least squares (PLS).

First, start with the SVD decomposition of the input matrix: $ X = U D W^T $. If $ n > p $ then of full rank. Here $ D = \mathrm{diag} ( d_1 , \ldots , d_p ) $ nonzero ordered $ e_1 > \ldots > e_p $ singular values. Then $W = ( w_1 , \ldots , w_p ) $ is the matrix of eigenvectors for $ S = X^T X $. 
We can then transform the original first layer to an 
orthogonal regression, namely $ y = ( UD ) W^T \beta $ with corresponding  OLS estimator $ \hat{\alpha} = ( Z^T Z )^{-1} Z^T y = D^{-1} U^T y $.

\paragraph{PLS and SVD Algorithm}
PCR has a long history in statistics. This is 
an unsupervised approach to dimension reduction (no $y$'s). 
Specifically, we first  center and standardize $ (y, \bm x )$.

Then, we provide an SVD  decomposition of 
$$ V: = \mathrm{ave} ( \bm x \bm x^T )  = \dfrac{1}{n}\sum_{i=1}^{n}\bm x_i\bm x_i^T$$  
This find the eigen-values $ e_j^2 $ and eigenvectors arranged in descending order, 
so we can write 
$$ V = \sum_{j=1}^p  e_j^2 {\bm v}_j {\bm v}_k^T .
$$ 
This leads to a sequence of regression models $(\hat Y_0, ..., \hat Y_K)$ with $ \hat Y_0 $ being the overall mean and 
$$
\hat{Y}_L = \sum_{l=0}^K (\mathrm{ave} ( w_l^T \bm x ) / e_l^2 ) \bm v_l^T \bm x 
$$
Therefore, PLS finds  ``features"  $Z_K = \{\bm v_k^T \bm x\}_{k=0}^K = \{\bm f_k \}_{k=0}^K $.

\paragraph{PCA and multivariate output} PCA emulation requires us to compute a reduction of multivariate output $Y$ using singular value decomposition of $Y$ by finding eigenvectors of $Z = \mathrm{ave}(YY^T)$. Then, we assume that the output is a linear combination of the singular vectors
\[
Y = w_1z_1 + \ldots,w_kz_k,
\]
where the weights $w_i$ follow a Gaussian Process. 
\[
z\sim \mathrm{GP}(m,K).
\]
Hence the method can be highly non-linear. This method is typically used when input variables come from a design of experiment. If interpretability of factors is not important,  and from a purely predictive point of view, PLS will lead to improved performance. 

One can view deep learning models as non-stochastic hierarchical data transformations. THe advantage is that we can learn deterministic data transformations before applying a stochastic model. That allows us to establish the connection between the Brillinger result and use of deep learning models and to develop a unified framework for modeling complex high-dimensional data sets. The prediction rule can be viewed as interpolation. 

In high-dimensional spaces you can mix-and-match the deterministic and stochastic data transformation rules. 

\paragraph{Model selection (a.k.a. dimension reduction)}
The goal of PCR is to minimize predictive MSE
$$
\hat{L} = \arg \min_K \mathrm{ave}( y - \hat{y}_K )^2 
$$
The choice of $K$ is determined via predictive  cross-validation. The $l$th model is simple regression of $ y $ on $ f_L = \bm v_l \bm x,\; l = 1 , \ldots, K $. 
\cite{mallows_comments_1973} $ C_p $ and $ C_L $ provide the relationship between shrinkage and model selection.

\paragraph{Dropout} This is a model selection technique designed 
to avoid over-fitting in deep learning.  This is done by  removing input dimensions in $X$ randomly with a given probability $p$. 
For example, suppose that we wish to minimize MSE,  $ \|Y-\hat{Y}\|^2_2$, then, when marginalizing over the randomness, we have a new objective
$$
{\rm arg \; min}_W \; \mathbb{E}_{ D \sim {\rm Ber} (p) } \Vert Y - W ( D \star X ) \Vert^2_2\,,
$$
This is equivalent to, with $ \Gamma = ( {\rm diag} ( X^\top X) )^{\frac{1}{2}} $, 
$$
{\rm arg \; min}_W \;  \Vert Y - p W X \Vert^2_2 + p(1-p) \Vert \Gamma W \Vert^2_2\,,
$$
Hence, this is equivalent to a Bayesian ridge regression with a $g$-prior as an objective function and reduces the likelihood of over-reliance on small sets of input data in training.

PLS provides three diagnostic plots: scree plot for dimensionality selection, by-plot and the correlation plot. The by-plot and the correlation plot allows the modeler to see how the output and input variables weight on hidden features. 

\subsection{Partial Least Squares} (PLS) transforms $Y,X \rightarrow U,T$  where $U$ are the $Y$ scores and $T$ are the $X$ scores respectively. By construction the loadings $P$ and $Q$ that correspond to regressions of $Y$ on $U$ and $X$ on $T$ are designed so that $U$ and $T$ have maximum correlation. So the prediction rule $\hat Y = Q\hat U = QTX$ will have the lowest possible in-sample MSE fit. THis contrasts with PCA which simply looks for variations of maximum explanation in the $X$ space without regard of the predictive ability of $Y$. PLS, therefore, provides the optimal patten matching data transformation methods\footnote[1]{There is no logical reason why the output variable needs to be closely related to the principal components, see \cite{cox68,polson2012}. Moreover, \cite{cook2007fisher} warned of pitfalls of using $Y$ for identifying dimensionality reduction transformations of $X$, in his analysis of agricultural field trials, the predictions should be chosen without output (the crop yield), thus making PLS inappropriate. Note, that requirement of $X$ being stochastic is not always satisfied. An important example is when $X$ is a result of design of experiment}. PLS also has a number of other advantages, which are specific to high dimensional problems. First, it handles multi-collinearity (as opposed to OLS). Second, it allows to handle multivariate $Y$. 

dimension reduction of $X$ depends on $Y$, the supervised learning reduction is more efficient.  The alternative non-linear un-supervised approach is to use an autoencoder. \cite{polson2021deep} show that one can avoid the linear assumption of PLS by combining with other non-linear techniques and improve the predictive power of a model.  

\cite{Ng2015ASE} provides a simple summary of PLS and shows that $P$ are singular vectors of $X^TY$

Partial Least Squares (PLS) uses both $X$ and $Y$ to calculate the projection, further it simultaneously finds projections for both input $x$ and output $y$, it make it applicable to the problems with high-dimensional output vector as well as input vector. 
Let $Y$ be $ n  \times q$ matrix of observed outputs and $ X $ be an $ n \times p$ input matrix. PLS finds a projection directions that maximize covariance between $X$ and $Y$, the resulting projections $U$ and $T$ for $Y$ and $X$, respectively are called  score matrices, and the projection matrices  $P$ and $Q$ are called loadings. The $X$-score matrix $T$, has $L$ columns, one for each ``feature" and $L$ is chosen via cross-validation. The key principle is that $T$ is a good predictor of $U$, the $Y$-scores. This relations are summarized by the equations below
\begin{align*}
Y &  =  U Q+ E \\
X & =  T P + F
\end{align*}
Here $Q$ and $P$ are orthogonal projection (loading) matrices and $T$ and $U$ are $n\times L$ are projections of $X$ and $Y$ respectively.

Originally PLS was developed to deal with the problem of collinearity in observed inputs. Although, principal component regression also addresses the problem of collinearity, it is often not clear which components to choose. The components that correspond to the larges singular values (explain the most variance in $X$) are not necessarily the best ones in the predictive settings. Also ridge regression addresses this problem and was criticized by  \citep{fearn_misuse_1983}. Further ridge regression does not naturally provide projected representations of inputs and outputs that would make it possible to combine it with other models as we propose in this paper. Thus, PLS seem to be the right method for high-dimensional problems when the goal is to model non-linear relations using another model. 

In the literature, there are two types of algorithms for finding the projections \citep{manne_analysis_1987}. The original one proposed by \cite{wold_collinearity_1984} which uses conjugate-gradient method \citep{golub2013matrix} to invert matrices. The first PLS projection $p$ and $q$  is found by maximizing the covariance between the $X$ and $Y$ scores 

\maxdisp{p,q}{\left(Xp\right)^T\left(Yq\right)}{||p||=||q||=1.} 

Then the corresponding scores are
\[
t = Xp,\text{   and   } u = Yq
\]
We can see from the definition that the directions (loadings) for $Y$ are the right singular vectors of $X^TY$ and loadings for $X$ are the left singular vectors. The next step is to perform regression of $T$ on $U$, namely $U = T\beta$.  The next column of the projection matrix $P$ is found by calculating the singular vectors of the residual matrices $(X - tp^T)^T(Y - T\beta q^T)$. The final regression problem is solved $Y = UQ = T\beta Q = XP^T\beta Q $. Thus the PLS estimate is
\[
\beta_{\mathrm{PLS}} = P^T\beta Q.
\]
\cite{helland_partial_1990} showed that PLS estimator can be calculated as 
\[
\beta_{\mathrm{PLS}} = R(R^TS_{xx}R)^{-1}R^TS_{xy}
\]
where $R = (S_{xy},S_{xx}S_{xy},\ldots,S_{xx}^{q-1}S_{xy})$,
\[
S_{xx} = \dfrac{X^T(I-{\bf 11^T}/n)X}{n-1},
\]
\[
S_{xy}  = ave ( y \bm x ). 
\]
\cite{helland_structure_1988} proposed an alternative algorithm to calculate the parameters 
For $K = 1 , \ldots, p$, set $ y_0 = y ,   \bm x_0 = \bm x $. Let  $ (y, \bm x )$ be centered and standardized. Given $ V = ave ( \bm x \bm x^T  ) $ and $ \bm s = ave ( y \bm x ) $. For $K= 1 $ to $p$ do 
\[
	\bm s_k = V^{K-1} \bm s, \quad \hat{y}_K = OLS ( y \; {\rm on} \; ( \bm s_k^T \bm x )_{k=1}^K ).
\]

\subsection{DL-PLS}
Partial least squares algorithm finds projections of the input and output vectors $X = TP+F$ and $Y = UQ+E$ in such a way that correlation between the projected input and output is maximized.  Our DL-PLS model will introduce nonlinearity $U  = G(T)$ by assuming that $U$ is a deep learner of $T$. From Brillinger's result we see that linear PLS calculates $T$ and $P$ for arbitrary $G_L$. 
\begin{align*}
Y &  =  U Q+ E \\
U & = G(T) \\
T & =  X P^T 
\end{align*}
where $G$ is a deep learner. 
Here $ X = T P + F $ is inverted to $ T = X P^T $ as $ P ^T P = I $.

Although we use a composite model DL-PLS, our estimation procedure is two-step. We first estimate the score matrices and then estimate parameters of the deep learning function $G$. This two step process is motivated by the \cite{brillinger_generalized_2012}. The results of Brillinger guarantee that  matrices $P, Q$ are invariant (up to proportionality) to nonlinearity, even when the true relationship between  $Y$-scores and $ X $-scores is nonlinear.

\paragraph{PLS-ReLU}
For example, $G$ is a simple feed-forward ReLU neural network, for which sparse Bayesian priors are useful to improve the generalization (\cite{polson2018posterior}). $U$ and $T$ are $n\times L$ matrices, 
\begin{align*}
	T &= Z_0\\
	Z_1 &= \max(Z_0W_1+b_1,0)\\
	U &= Z_1W_2+b_2
\end{align*}
The weights $W_1$ and $W_2$ are to be learned.

\paragraph{PLS-CNN}
Partial least squares can also be used as a layer at any stage of deep learning. For example, in a convolutional neural network, 
$$
Z_1 = g\left(\sum_{i\in M} X_i\star w_i + b_i\right)
$$
where $X\star w + b$ denotes the convolution over the region $M$, with input image $X$, weights $w$ and bias $b$. Then we can add a PLS layer by regressing the output $Y$ on $Z_1$ and perform feature reduction, which results in a CNN-PLS model.

\paragraph{PLS-Autoencoder}
As the input $T$ and output $U$ have the same dimensions, one can consider using an autoencoder network. We build on the architecture of \cite{zhang2019improved}. 
\begin{align*}
	T &= Z_0\\
	Z_1 &= g_1(Z_0W_1+b_1)\\
	U &= g_2(Z_1W_2+b_2)
\end{align*}
where $Z_1$ is a $n\times l_1$ matrix which acts as a lower dimensional intermediate hidden layer and $l_1 \ll l_0 = L$.


\paragraph{DL-RNN}
The idea of using a sequence of transformations was also previously considered for the analysis of temporal data \cite{wiener1964extrapolation,singpurwalla2018least,west1981robust}. Work of \cite{masreliez1975approximate} provides an equivalent decomposition of Brillinger for time series models. 

The deep learning analog is the recurrent architecture that uses autoregression in the latent feature space. Let $Y_t$ denote the observed response and $Z_t$ are hidden states, then the RNN model is:

\begin{align*}
Y_t \mid Z_t & \sim P(Y_t \mid z_t ),\\
Z_t = & W_za_t + b_{2}\\
a_{t} & = f_1(W_{1}[Z_{t-1}, X_{t}] + b^{1})
\end{align*}
where $f_1$ is an activation function such as $\tanh(x)$. The time invariant weight matrices $W_1$ and $W_{z}$ are found through training the network. $X_t$ are external inputs up to $k$ lags, $Z_{t-1}$ are the previous hidden states, and the hidden state is initialized to zero, $Z_{t-k}=0$.

The main difference between RNNs and feed-forward deep learning is the use of a hidden layer with an auto-regressive component, here $W^{1}_{z}Z_{t-1}$. It leads to a network topology in which each layer represents a time step, indexed by $t$, in order to highlight the temporal nature. 
\begin{figure}[H]
\begin{center}
\includegraphics[width=0.28\linewidth]{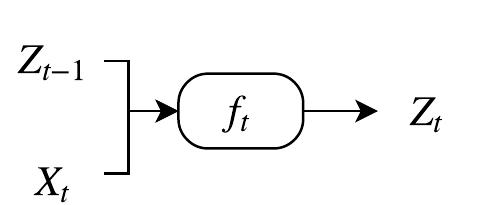}
\end{center}
\caption{Hidden layer of a Recurrent Neural Network.}
\label{fig:hl-rnn}
\end{figure}
Additional depth can be added to create deep RNNs by stacking layers on top of each other, using the hidden state of the RNN as the input to the next layer. RNNs architectures are learned through the same mechanism described for feedforward architectures. One key difference between implementations of RNNs is that drop-out is not applied to the recurrent connections, only to the non-recurrent connections. In contrast, drop-out is applied to all connections in a feedforward architectures.

\paragraph{DL-PLS-GP}
Given a new predictor matrix $X_{*}$ of size $N_*\times p$, the same projection $P$ produces the corresponding $F_{*} = [f_{1,*}, ..., f_{L, *}]$. We can use Gaussian process regression to predict $U_{*} = [u_{1,*}, ..., u_{L, *}]$ from $F_{*}$ as follows
\begin{align*}
F_{*} &= X_{*} P^T\\
\hat u_{k, *} &= g_{k,*}(f_{k, *}), k=1,2,...,L.
\end{align*}
where $g_{k,*}$'s are the Gaussian process regression predictors. 

\begin{equation*}
\begin{bmatrix}
u\\
u_*
\end{bmatrix}\sim N\left(
\begin{bmatrix}
g\\
g_*
\end{bmatrix},
\begin{bmatrix}
K & K_*\\
K_*^T & K_{**}
\end{bmatrix}\right)
\end{equation*}
where $K = K(t, t)$ is $N\times N$, $K_* = K(t, t_*)$ is $N\times N_*$, and $K_{**} = K(t_*,t_*)$ is $N_*\times N_*$. $K(\cdot,\cdot)$ is a kernel function. The conditional mean, $g_*$, is given by
\begin{align*}
g_*(t_*) &= g(t_*) + K_*^TK^{-1}(u - g(t)). 
\end{align*}
Then prediction of $Y$ is
\begin{equation*}
\hat Y_* = \hat U_{*} Q
\end{equation*} 

\section{Finding Good Prediction Rules}\label{sec:predictors}
There are two ways of finding good prediction rules in high dimensions. One is via shrinkage and the other is based on ensemble $1/N$ rules. 
\subsection{Shrinkage}
We will rely heavily on the Bayesian shrinkage interpretation of PCR and PLS due to Frank and Friedman (1993). \cite{polson2010,polson2012} provide a general theory of global-local shrinkage and, in particular,  analyze $g$-prior and horseshoe shrinkage. PLS behaves differently from standard shrinkage rules as it can shrink away from the origin for certain eigen-directions.

The corresponding shrinkage factors for RR and PCR are typically normalized so that they give the same overall shrinkage so that the length of solution vector are the same ($|\hat\beta_{RR}| = |\hat\beta_{PLS}|$).

This scale factors provide a diagnostic plot: $f_j $.  If any $ f_j > 1 $ then one can expect 
supervised learning (a.k.a. PLS with $Y$'s influence the scaling factors) will lead to different  predictions than   unsupervised learning (a.k.a. PCR with
solely dependent  on $X$).  In this  sense, PLS is an optimistic procedure in that the goal is  
to maximise the explained variability of the output in sample with the  hope of generalizing well out-of-sample. For linear estimator, $f_j>1$ means that both the bias and the variance are increased. \cite{frank_statistical_1993} mention the possibility of improving the performance of PLS by modifying the scale factors as $\tilde f_j^{PLS} \leftarrow \min\left\{f_j^{PLS}, 1\right\}$, although it's not certain since PLS is not linear. The shrinkage factors of PLS are also discussed in \cite{rosipal2005overview} and are closely related to the Ritz pairs.

The key insight is that all of the estimators are of the from 
$$ 
\hat Y^M = \sum_{j=1}^L  f_j^M  \hat{\alpha}_j \bm v_j^T \bm x 
$$
where $f_j$ are scale factors. $M$ denotes method (e.g. RR, PCR, PLS). $L$ is the rank of $\bm V$ (number of nonzero $e_k^2$). 
For PCR, the scale factors are  $f_j =1 $  for top $L$ eigenvectors  \cite{frank_statistical_1993} 
\begin{align*}
f_j^{RR} &= e_j^2/(e_j^2 + \lambda ), \text{where $\lambda$ is a fixed regularization parameter}\\
f_j^{PCR} &= \begin{cases}
1, & e_j^2 \geq e_L^2\\
0, & \text{otherwise}
\end{cases}\\
f_j^{PLS} &= \sum_{k=1}^{K} \theta_k e_j^{2k} \; ,  \; {\rm where} \;  \theta = w^{-1}\eta ,  \;  \eta_k = \sum_{j=1}^p \hat{\alpha}_j^2 e_j^{2(k+1)} . 
\end{align*}

The Bayesian paradigm provides novel insights into how to construct estimators with good predictive performance. The goal is simply to find a good predictive MSE, namely $E_{Y,\hat{Y}}(\Vert\hat{Y} - Y \Vert^2)$, where $\hat{Y}$ denotes a prediction value. Stein shrinkage (a.k.a regularization with an $L^2$ norm) in known to provide good mean squared error properties in estimation, namely $E(||\hat{\theta} - \theta)||^2)$. These gains translate into predictive performance  (in an iid setting) for $E(||\hat{Y}-Y||^2)$. 

The main issue is how to tune the amount of regularization (a.k.a prior hyper-parameters).  Stein's unbiased estimator of risk provides a simple empirical rule to address this problem as does cross-validation. From a Bayes perspective, the marginal likelihood (and full marginal posterior) provides a natural method for hyper-parameter tuning. The issue is computational tractability and scalability. In the context of DL, the posterior for $(W,b)$ is  extremely high dimensional and multimodal and posterior MAP provides good predictors $\hat{Y}(X)$. 

\subsection{Ensemble Predictors}
\begin{figure}[H]
	\centering
\begin{tabular}{cc}
	\includegraphics[width=0.5\textwidth]{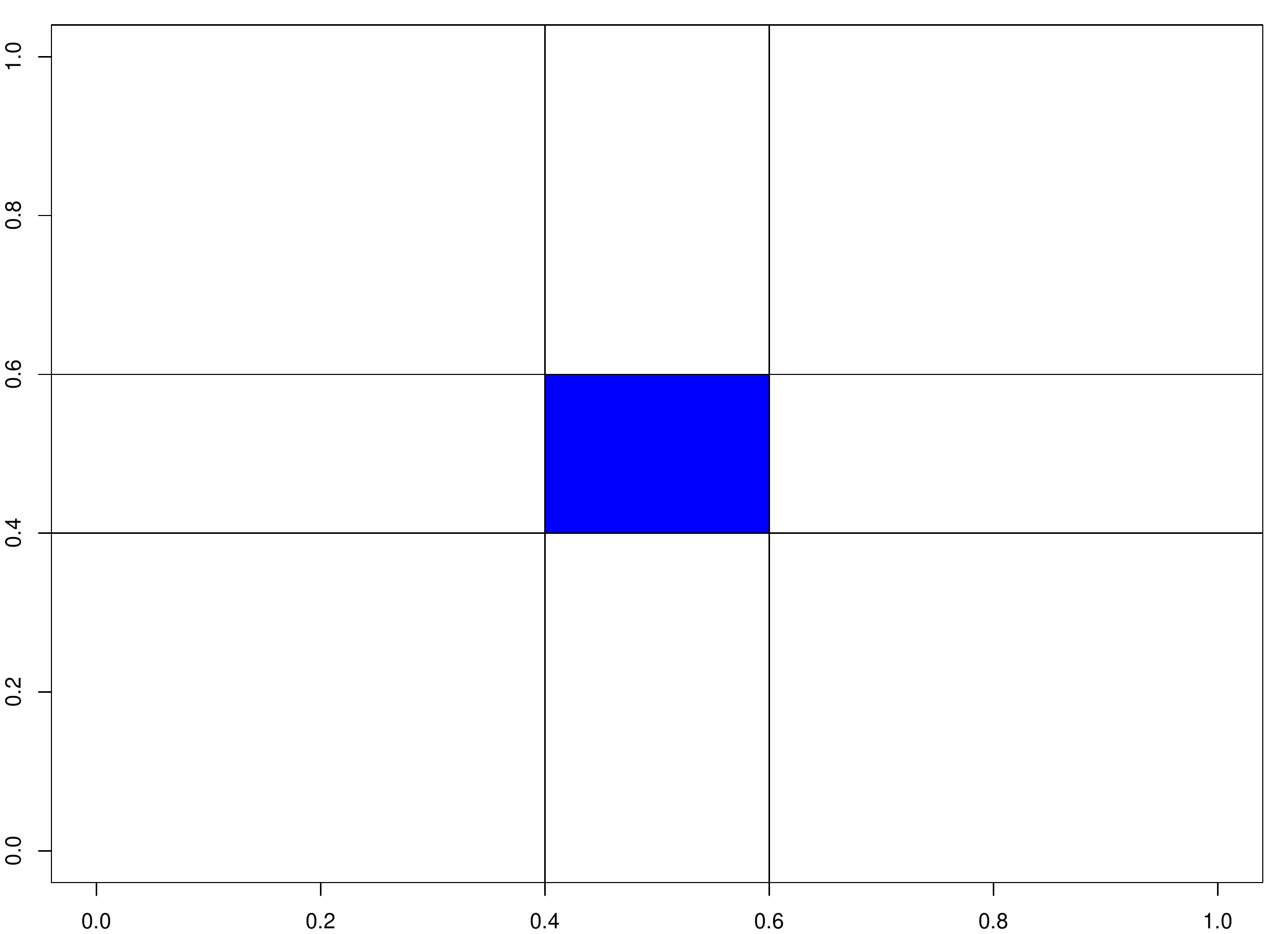}& 	\includegraphics[width=0.5\textwidth]{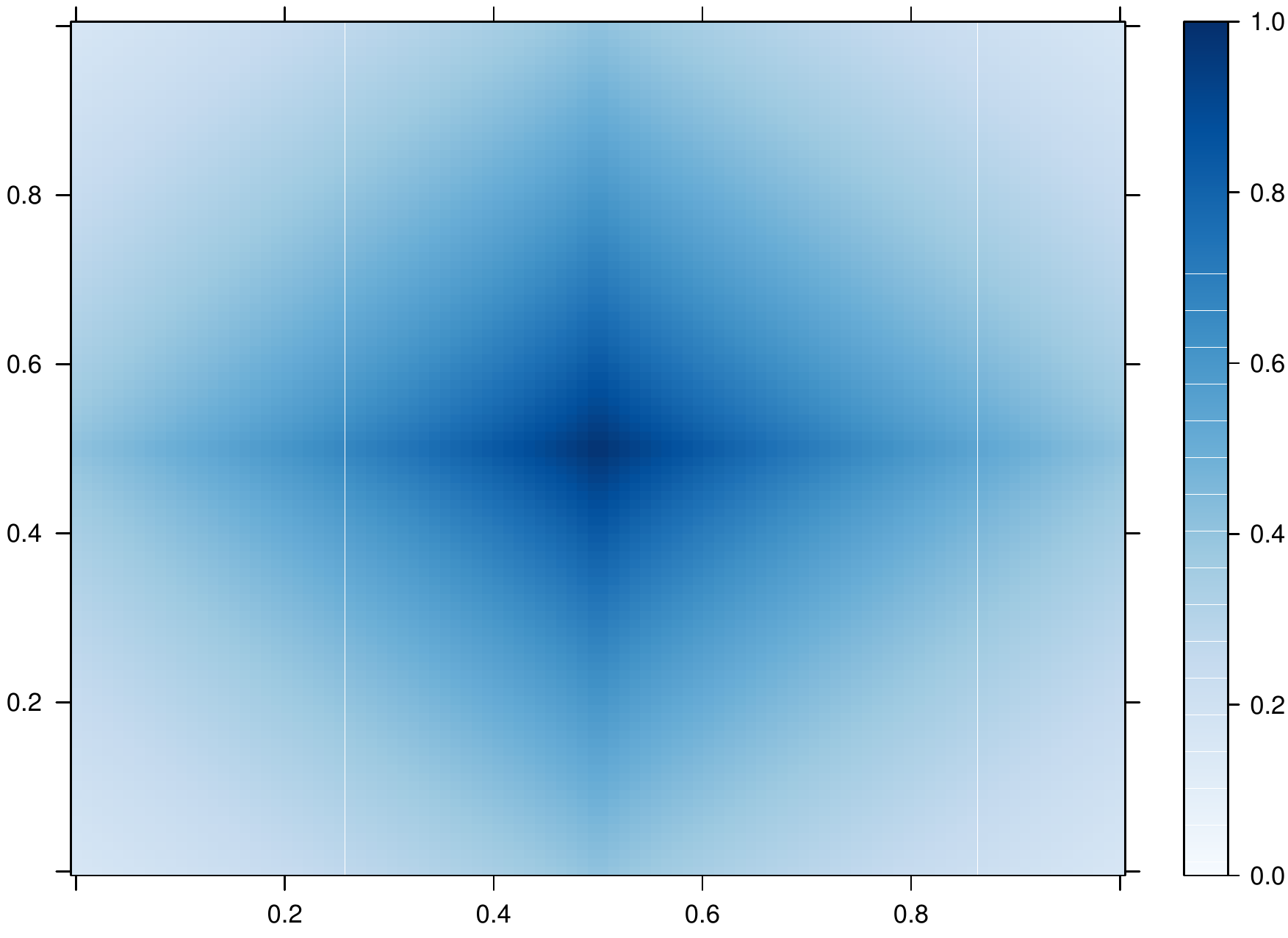}\\
	(a) Tree Kernel & (b) Random Forest Kernel
\end{tabular}
\caption{Kernel Weight. The intensity of the color is proportional to the size of the weight. Left panel (a) shows weights for tree-based model, with non-zero values only inside a cylindrical region (a box), and (b)  shows weights for a random forest model, with non-zero wights everywhere in the domain and sizes decaying away from the location of the new observation.}
\label{fig:cilinder}
\end{figure}

Bayes conditional averaging performs well in high dimensional regression and classification problems. High dimensionality, however, brings with it the curse of dimensionality and  it is instructive to understand why certain kernel can perform badly. Adaptive Kernel predictors (a.k.a. smart conditional averager) are of the form 
$$
\hat{Y}(X) = \sum_{r=1}^R K_r ( X_i , X ) \hat{Y}_r (X).
$$
Here $ \hat{Y}_r(X) $ is a deep predictor with its own trained parameters. For tree models, the kernel $ K_r( X_i , X) $ is a \emph{cylindrical} region $ R_r $ (open box set). Figure \ref{fig:cilinder} illustrates the implied kernels for trees (cylindrical sets) and random forests. Not too many points will be neighbors in a high dimensional input space.  

Constructing the regions to preform conditional averaging is fundamental to reduce the curse of dimensionality. Deep learning can improve on traditional methods by performing a sequence of GLM-like transformations. Effectively DL learns a distributed partition of the input space. 
For example, suppose that we have $K$ partitions and a DL predictor that takes the form of a weighted average or soft-max of the weighted average for classification.  Given a new high dimensional input $X_{\mathrm{new}}$, many deep learners are then an average of learners obtained by our hyper-plane decomposition. Our predictor takes the form
\[
\hat{Y}(X) = \sum_{k \in K} w_k(X)\hat{Y}_k(X),
\] 
where $w_k$ are the weights learned in region $K$, and $k$ is an indicator of the region with appropriate weighting given the training data.

\begin{figure}[H]
	\centering
\includegraphics[width=0.5\textwidth]{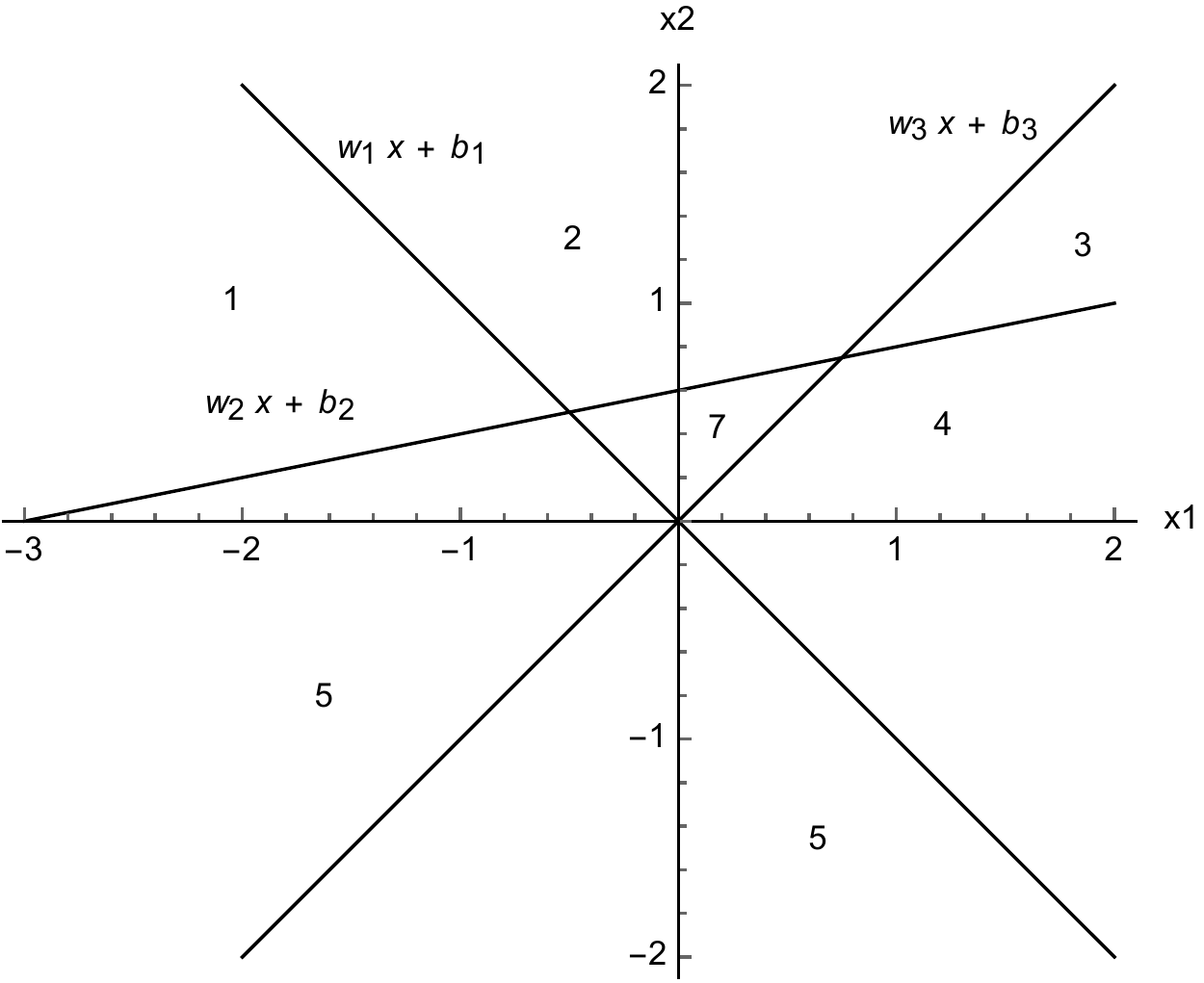}
\caption{Hyperplanes defined by three neurons with ReLU activation functions.}
\label{fig:hyper_planes}
\end{figure}

The partitioning of the input space by a deep learner is similar to the
one performed by decision trees and partition-based models such as
CART, MARS, RandomForests, BART, and Gaussian Processes. Each neuron in
a deep learning model corresponds to a manifold that divides the input
space. In the case of ReLU activation function $f(x) = \max(x,0)$ the
manifold is simply a hyperplane and the neuron gets activated when the new
observation is on the ``right'' side of this hyperplane, the activation
amount is equal to how far from the boundary the given point is. For
example in two dimensions, three neurons with ReLU activation functions
will divide the space into seven regions, as shown on Figure \ref
{fig:hyper_planes}.

The key difference between tree-based architecture and neural network based models is the way hyper-planes are combined. Thus, the number of hyper-planes grow exponentially with the number of layers. The key property of an activation function (link) is $f(0) = 0$ and it has zero value in certain regions. For example, hinge or rectified learner $\max(x,0)$  box car (differences in Heaviside) functions are very common. As compared to  a logistic regression, rather than using $\mathrm{softmax}(1/(1+e^{-x}))$ in deep learning $\tanh(x)$ is typically used for training, as $\tanh(0)=0$.

\cite{amit_shape_1997} provide an interesting discussion of efficiency. Formally, a Bayesian probabilistic approach (if computationally feasible) optimally weights predictors via model averaging with  $\hat{Y}_k(x) = E(Y \mid X_k)$
$$
\hat{Y}(X) = \sum_{r=1}^R w_k \hat{Y}_k(X).
$$

Such rules can achieve optimal out-of-sample performance.  \cite{amit2000multiple} discusses the striking success of multiple randomized classifiers. Using a simple set of binary local features, one classification tree can achieve 5\% error on the NIST data base with 100,000 training data points. On the other hand, 100 trees, trained under one hour, when aggregated, yield an error rate under 7\%. This stems from the fact that a sample from a very rich and diverse set of classifiers produces, on average, weakly dependent classifiers conditional on class. 

\paragraph{1/$N$ Ensamble Rules}
To further exploit this, consider the  model of weak dependence, namely exchangeability. This often occurs in high dimensional spaces, where it is easy to find a prediction rule with high variance and the $1/N$ rule by a portfolio argument reduces the variance. 

Suppose that we have  $N$ exchangeable, $ \mathbb{E} ( \hat{Y}_i ) = \mathbb{E} ( \hat{Y}_{\pi(i)} ) $, and stacked predictors
$$
\hat{Y} = ( \hat{Y}_1 , \ldots , \hat{Y}_N ). 
$$
Suppose that we wish to find weights, $w$, to attain $ {\rm arg \; min}_W \; E l( Y , w^T \hat{Y} ) $ where $l$ convex in the second argument;
\[
E l( Y , w^T \hat{Y} )  = \frac{1}{N!} \sum_\pi E l( Y , w^T \hat{Y} )
 \geq  E l \left ( Y , \frac{1}{N!} \sum_\pi w_\pi^T \hat{Y} )\right ) =  E l \left ( Y , (1/N) \iota^T \hat{Y} \right ) 
\]
where $ \iota = ( 1 , \ldots ,1 ) $. Hence, the randomized multiple predictor with weights $w = (1/N)\iota$  provides the optimal Bayes predictive performance. 

\subsection{Brillinger Estimation}
\cite{brillinger_generalized_2012} considers the single-index model with non-Gaussian regressors where  $(Y,X) $ are stochastic with conditional distribution
$$
Y \mid X \sim  N(g ( \alpha + \beta X ),~\sigma^2).
$$
Here $\beta  X  $ is the single features found by data reduction from high dimensional $ X$. 
Let $ \hat{\beta}_{OLS} $ denote the least squares estimator which solved $X^TY = X^TX\beta$.  By Stein's lemma, 
$$
\mathrm{cov}(Y,X)  = \beta \mathrm{cov} ( g ( \alpha + \beta X ) , \alpha + \beta X ) \mathrm{var}(X) / \mathrm{var}( \alpha + \beta X ) 
$$
Then $ \hat{\beta} $ is consistent as
$$
\hat{\beta}_{OLS} = \mathrm{cov}v( Y,X) / \mathrm{var}(X) \rightarrow k \beta \;  \; {\rm where} \; \; k = \cov ( g ( \alpha + \beta X ) , \alpha + \beta X ) \mathrm{var}(X) / \mathrm{var}( \alpha + \beta X ) 
$$
Hence,  $ \hat{\beta}_{OLS} $ estimator is proportional to $ \beta $. We can also non-parametrically estimate $g(u)$ by plotting 
$ ( \hat{\beta} x_j , y_j  ) , ~j = 1 , \ldots n $ and smoothing $y_j$ values with $  \hat{\beta} x_j $ near $ u $. 

Hence, when the $X$s are Gaussians  and independent of the error, we have the relationship $ \cov(Y,X) = k \beta \mathrm{var}(X) $. This relationship follows from the weaker assumption
$$
E( Y|X ) = g ( \alpha + \beta X ) 
$$
Hence, this approach can be applied to binary classification with $ Prob ( Y=1 | X ) = g( \alpha + \beta X ) $ and other models such as survival models.

If we substitute $F$ in equation \ref{eq:brf} with $g(Z)$ to get $p(Y \mid g(Z))$, to minimize predictive MSE for future $Y$ observations, we simply use the conditional mean $\hat Y = E\left(Y \mid g(Z)\right)$. In terms of the loading matrix $P$ we are to have the predictive rule 
\[
Y \mid X = E\left(Y \mid g(X^TP) \right).
\]
The Brillinger result $E(Y \mid g(z)) = G(X^TP)$  allows us to estimate the $P$ using PLS. Essentially, we have a multivariate index model. As long as we have stochastic regressions, we can identify the $P$ matrix. 
Surprisingly you can identify the weights $P$ and the latent variables  $ Z = X P $ that then determine the features $ F + X P $  using only OLS or PLS.

\section{Applications}\label{sec:applications}
\subsection{Identifying Nonlinear Systems}
To illustrate our methodology, we start with multivariate observation and state identification problem. We start with a simple linear example to show how sequential steps of PLS is used to estimate the coefficient matrix $P$.  
$$
Y = XP + \epsilon
$$
where $X \in \mathbb{R}^{N\times 3}$, $P \in \mathbb{R}^{3\times 2}$ and $Y \in \mathbb{R}^{N\times 2}$. $\epsilon$ is the matrix of $N$ i.i.d. noises drawn from normal distribution.  

Since it is a simple linear system, the OLS directly identifies $P$, however, it is instructive to see how the sequential PLS performs the same task. 

\begin{figure}
	\centering
	\includegraphics[width=0.3\linewidth]{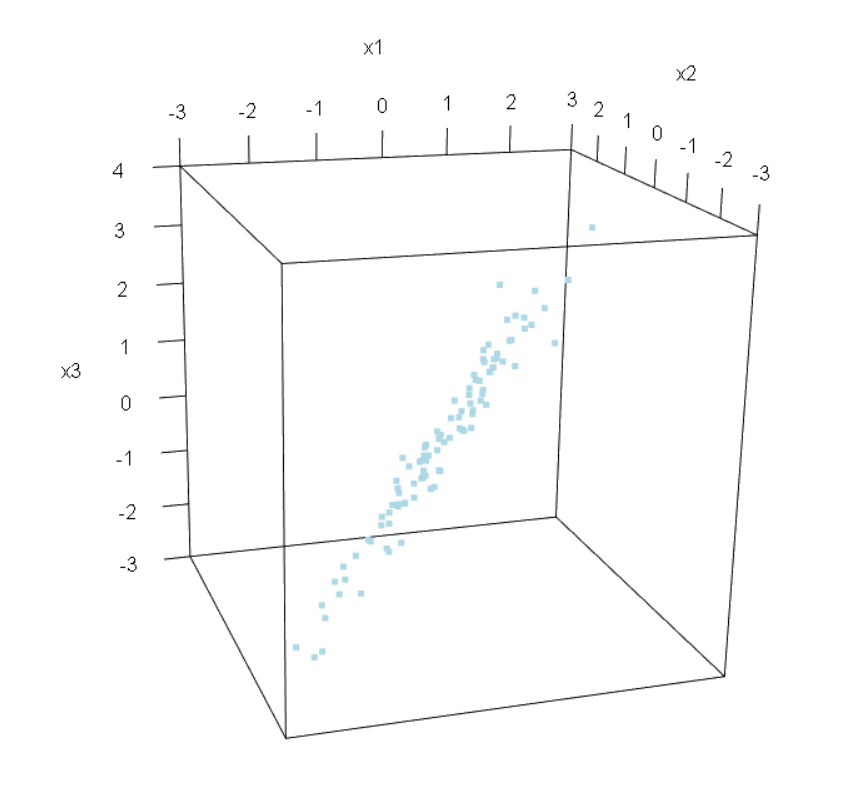}
	\includegraphics[width=0.3\linewidth]{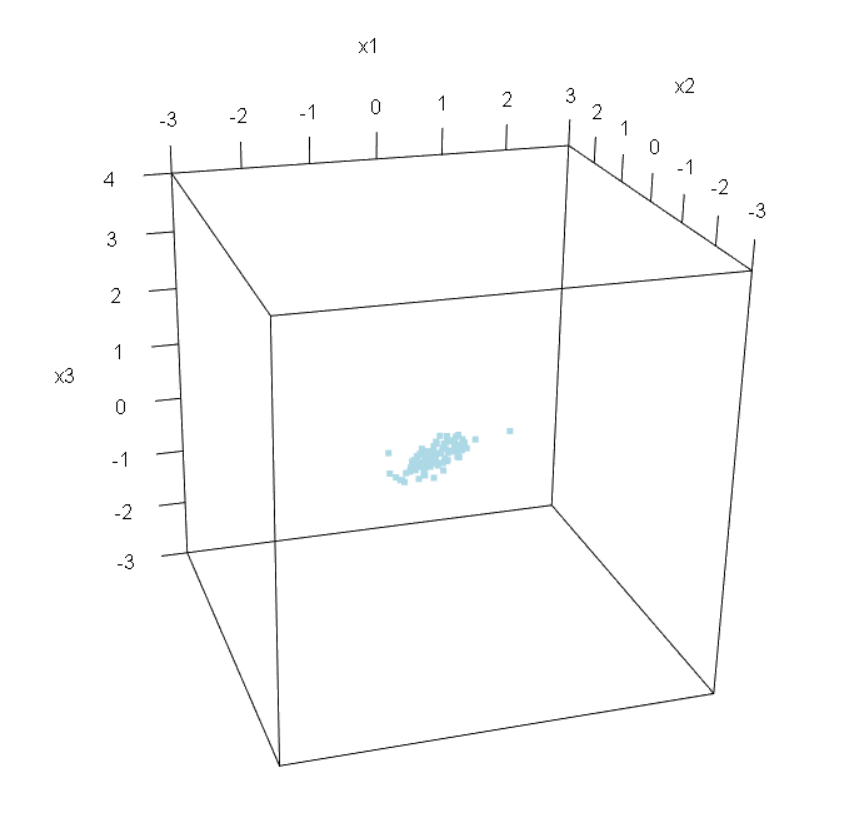}
	\includegraphics[width=0.3\linewidth]{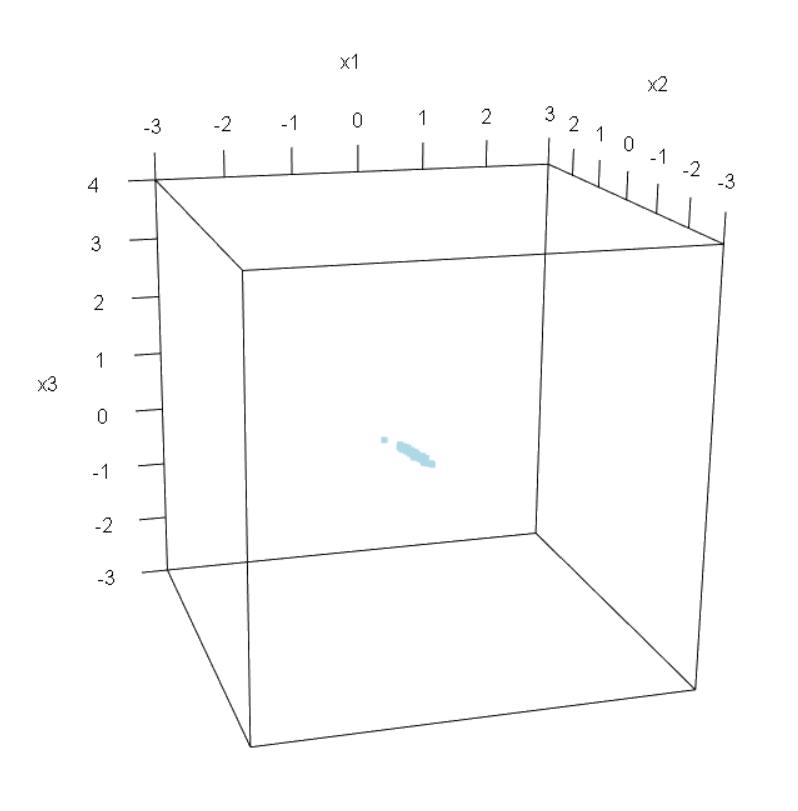}
		\includegraphics[width=0.3\linewidth]{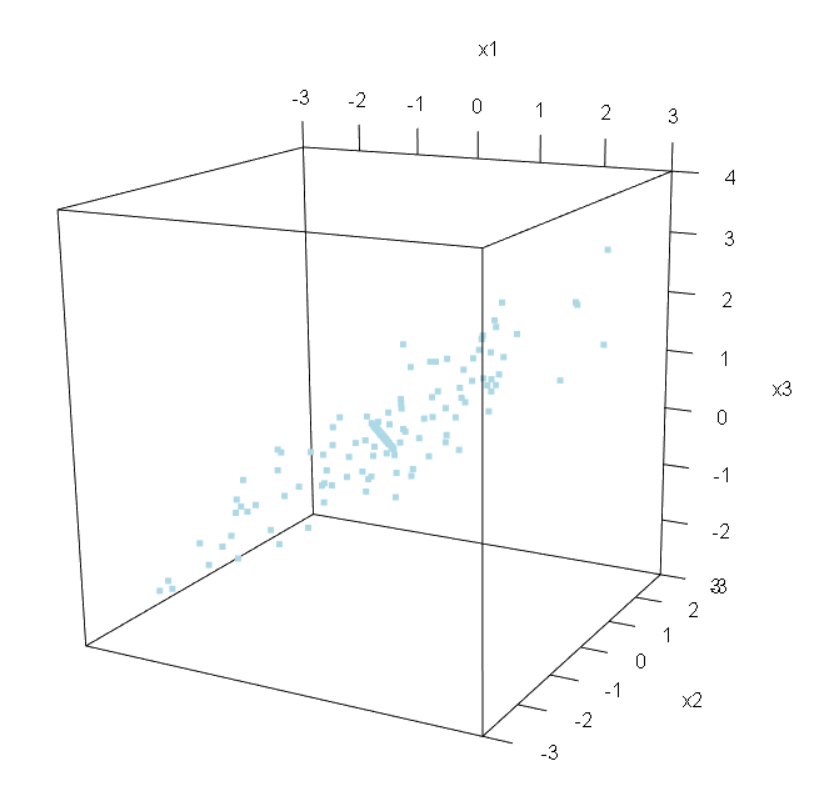}
	\includegraphics[width=0.3\linewidth]{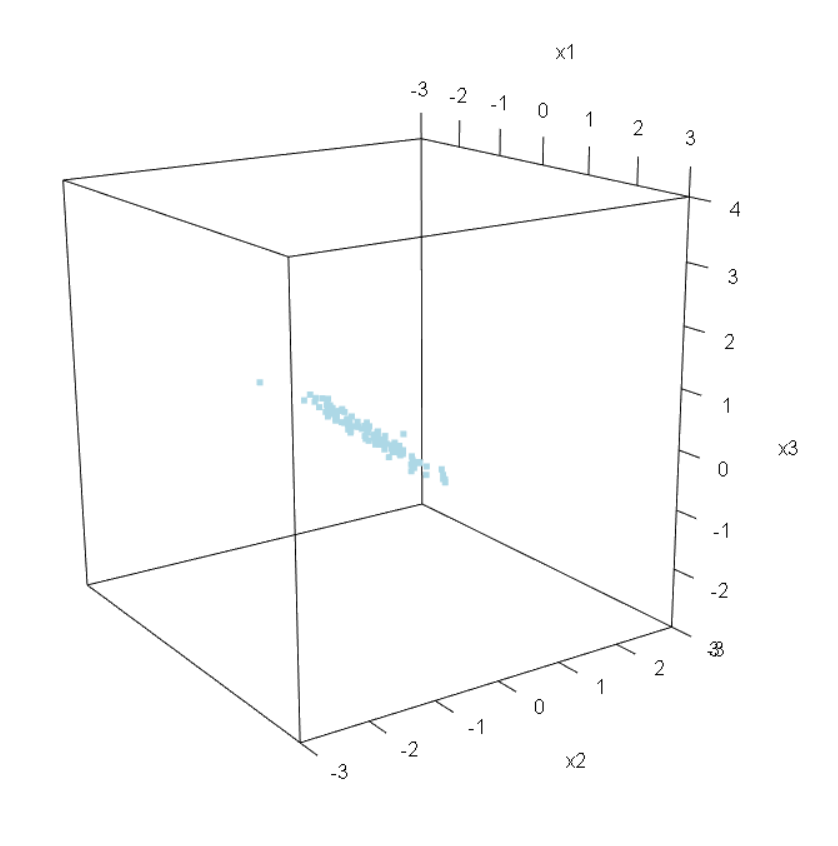}
	\includegraphics[width=0.3\linewidth]{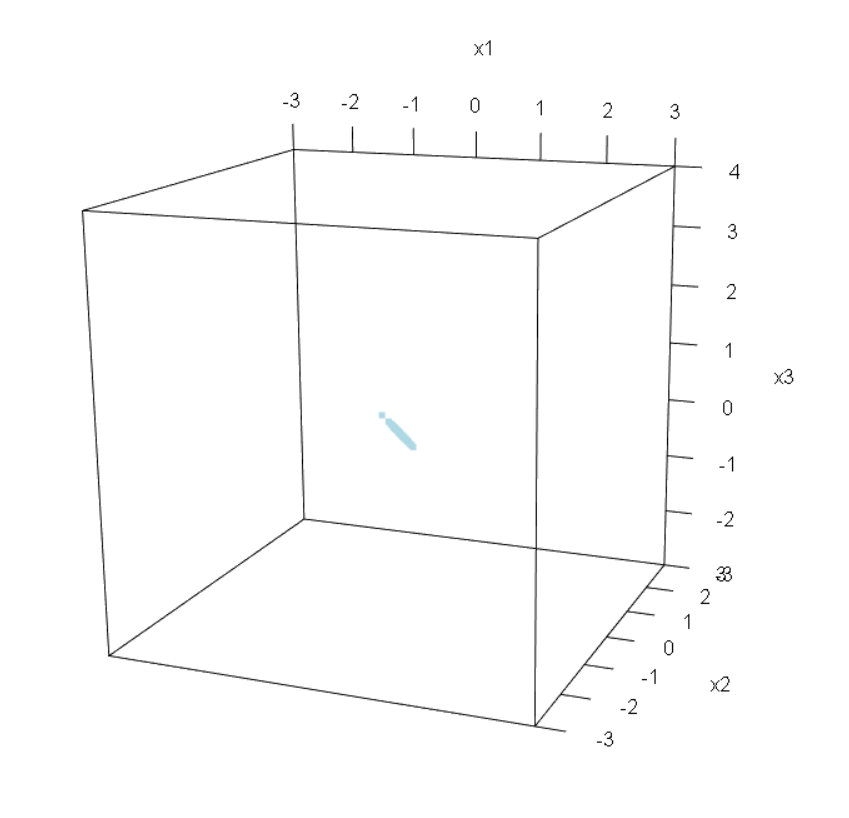}
	\caption{Input: each iteration extracts one component out from the data.  (from left to right: 1st round, 2nd round, third round.)}
	\label{fig:asada}
\end{figure}

Figure \ref{fig:asada} provides two rotational views of 2-dimensional $x$ variables and as we move from the left to right, the sequential implementation of the PLS algorithms to find the latent score features. Each iteration extracts one additional component of the data.

\begin{figure}
	\centering
	\includegraphics[width=0.9\linewidth]{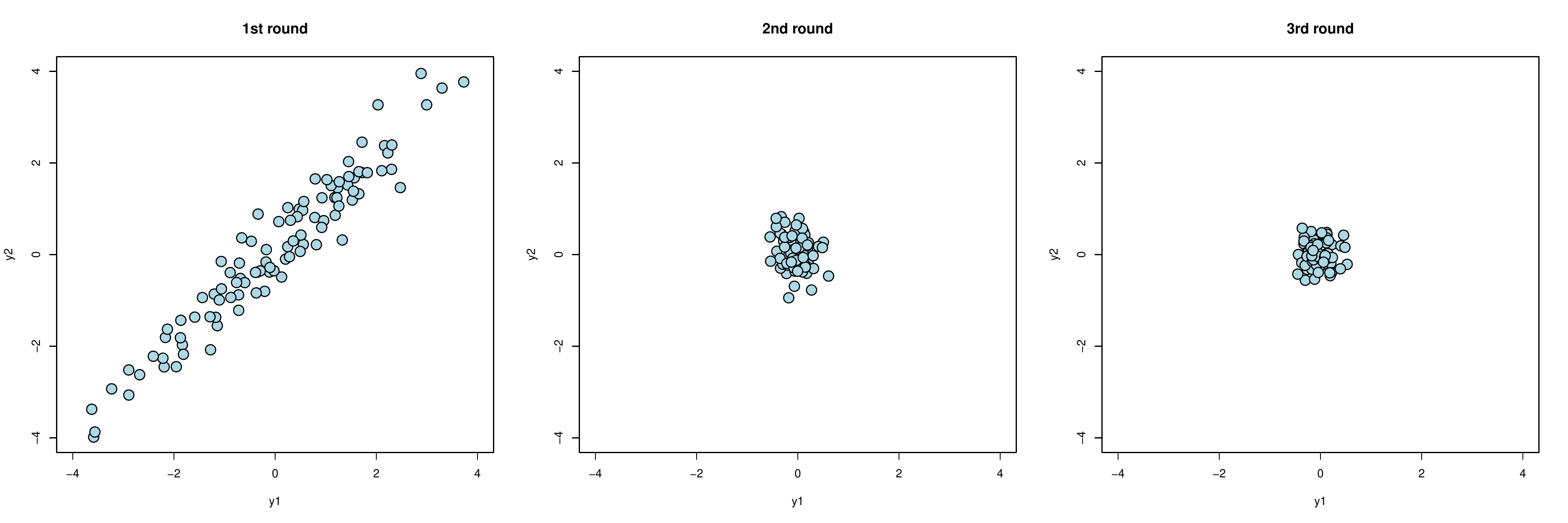}
	\caption{Output: all meaningful information is extracted after 2 iterations, leaving only noise.}
	\label{fig:asaday}
\end{figure}

Figure \ref{fig:asaday} shows the effect of the PLS data transformations on the output variable. To identify the system the goal is to transform the output variable until it looks like white nose. As you see in Figure \ref{fig:asaday} it occurs in one step and our dimensionality reduction is performed. The second step provides a marginal improvement. 
 
Correspondingly, when we regress the the multivariate output $y$ on $T$ scores, we recover the loadings matrix $P\ in R^{3 \times 2}$. Moreover, we can generalize it to a non-linear system using Brillinger result. We can use either OLS or PLS to identify the system. 

$$
Y =  | 10 + XP | + \epsilon  
$$
where $X\in\mathbb{R}^{N\times 100}$ and $P\in\mathbb{R}^{100\times 1}$. Using OLS or PLS, we can estimate the coefficient $P$ and successfully recover the nonlinear absolute function.

\begin{figure}
	\centering
	\includegraphics[width=0.9\linewidth]{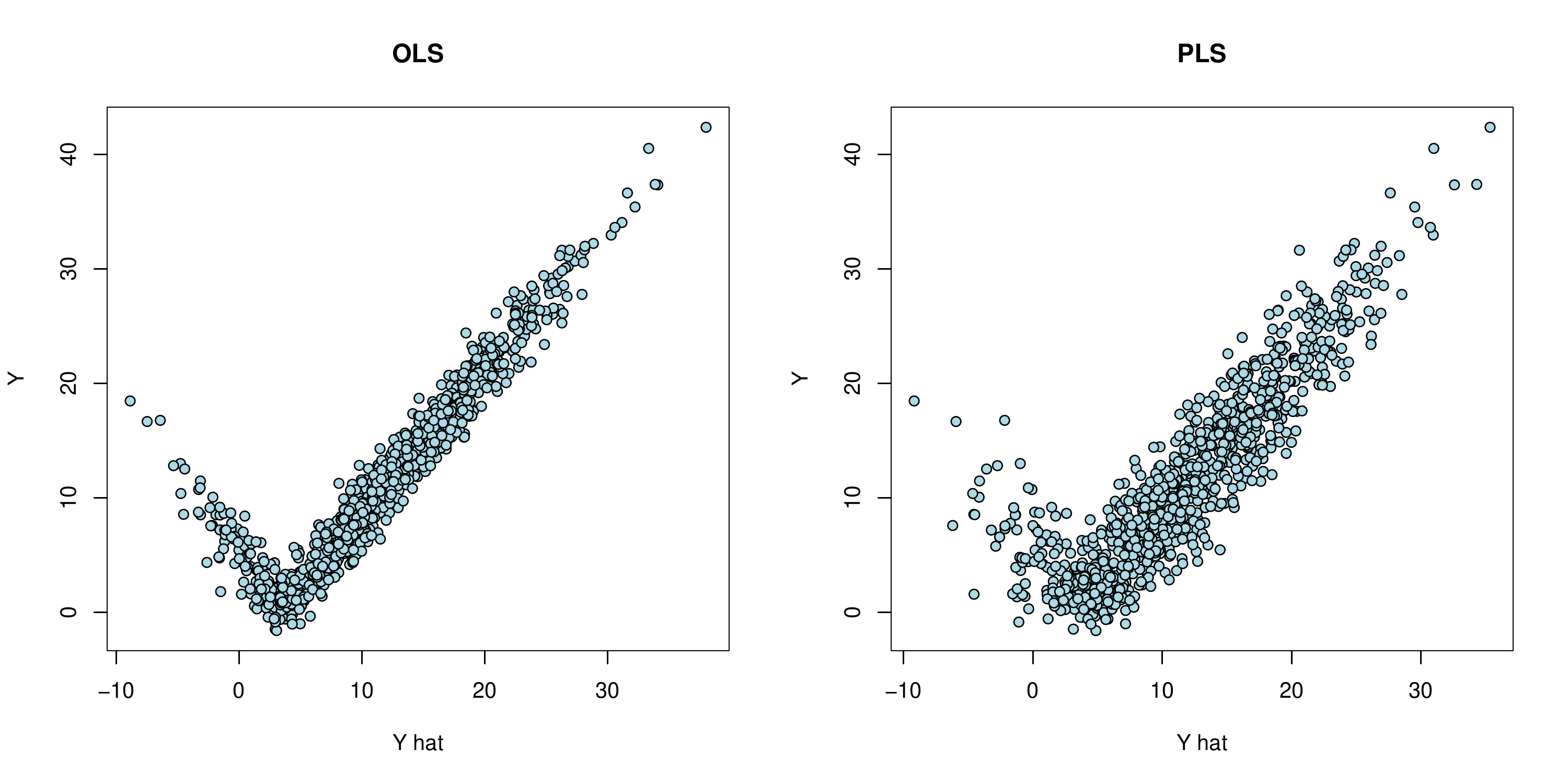}
	\caption{Nonlinear system: OLS and PLS estimators recover the nonlinear absolute function.}
	\label{}
\end{figure}

\subsection{Simulated Examples} 
\subsubsection{Dimension Reduction}
First, we consider how deep learning can be used as a sufficient dimensionality reduction technique to identify lower dimensional features that can be later used as inputs to statistical models. We start with a synthetic example that demonstrates an application of one layer neural network to find a piece-wise linear structure in the data. Consider real-valued function $f(x_0,\dots,x_{100})$,
\[
f(x) = |u^Tx|
\]
We generate input-output pairs $\{x_i,y_i\}_{i=1}^n$, where $x_{ij} \sim Unif[-1,1]$, and $y_i = f(x_i)+\epsilon,~~\epsilon\sim N(0,0.01)$. There is one-dimensional structure in the input-output relations which is represented by a ridge function $|u^Tx|$. We use neural network to identify this one dimensional structure. We introduce a one-dimensional bottleneck in our neural network. The overall architecture is as follows $\hat y  = F(x) = f(\phi(x))$, where $\phi: \mathbb{R}^p \rightarrow \mathbb{R}$, and $f: \mathbb{R} \rightarrow \mathbb{R}$. Both $\phi$ and $f$ are single layer neural networks. 
\begin{figure}[H]
	\begin{tabular}{cc}
		\includegraphics[width=0.5\linewidth]{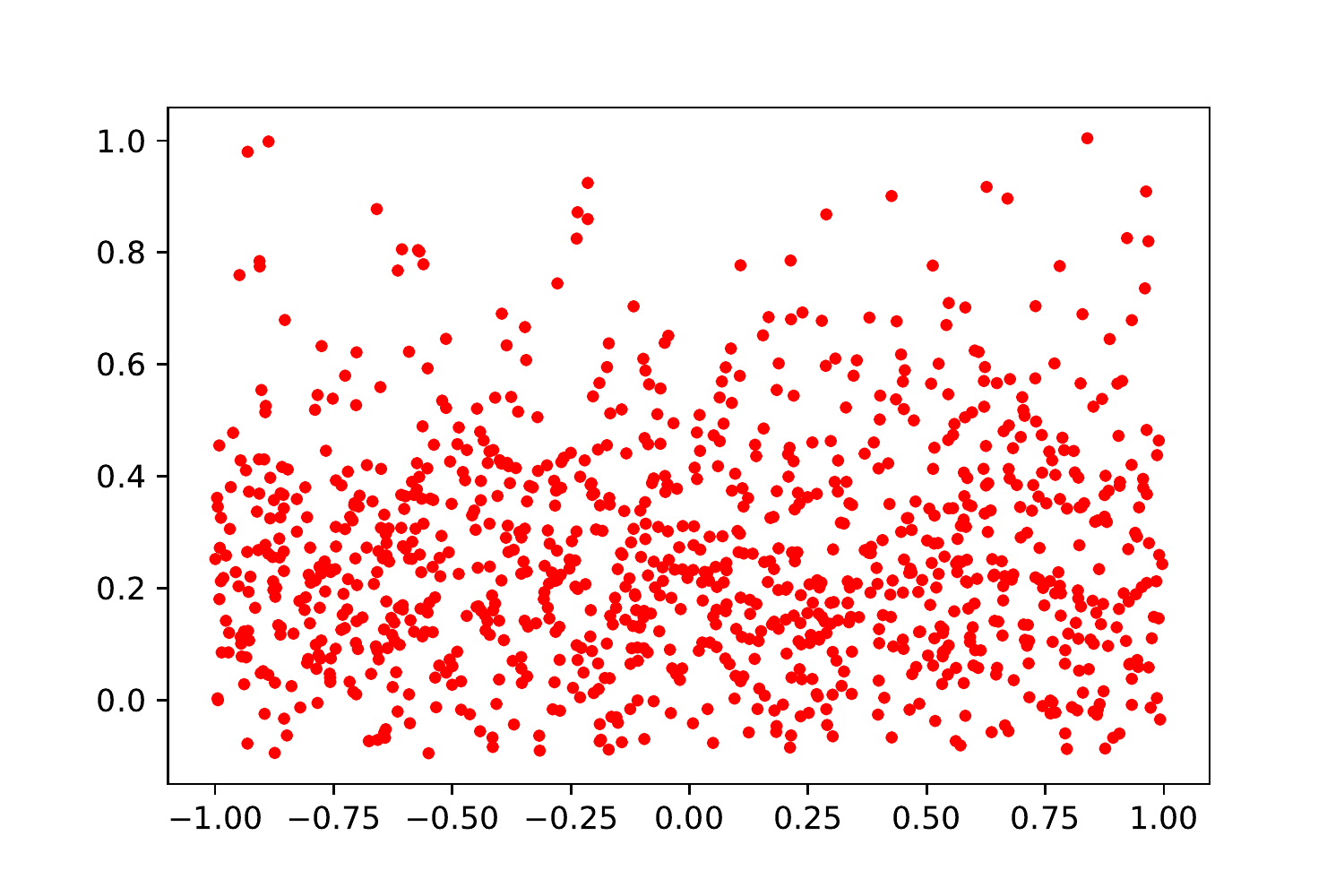}&
		\includegraphics[width=0.5\linewidth]{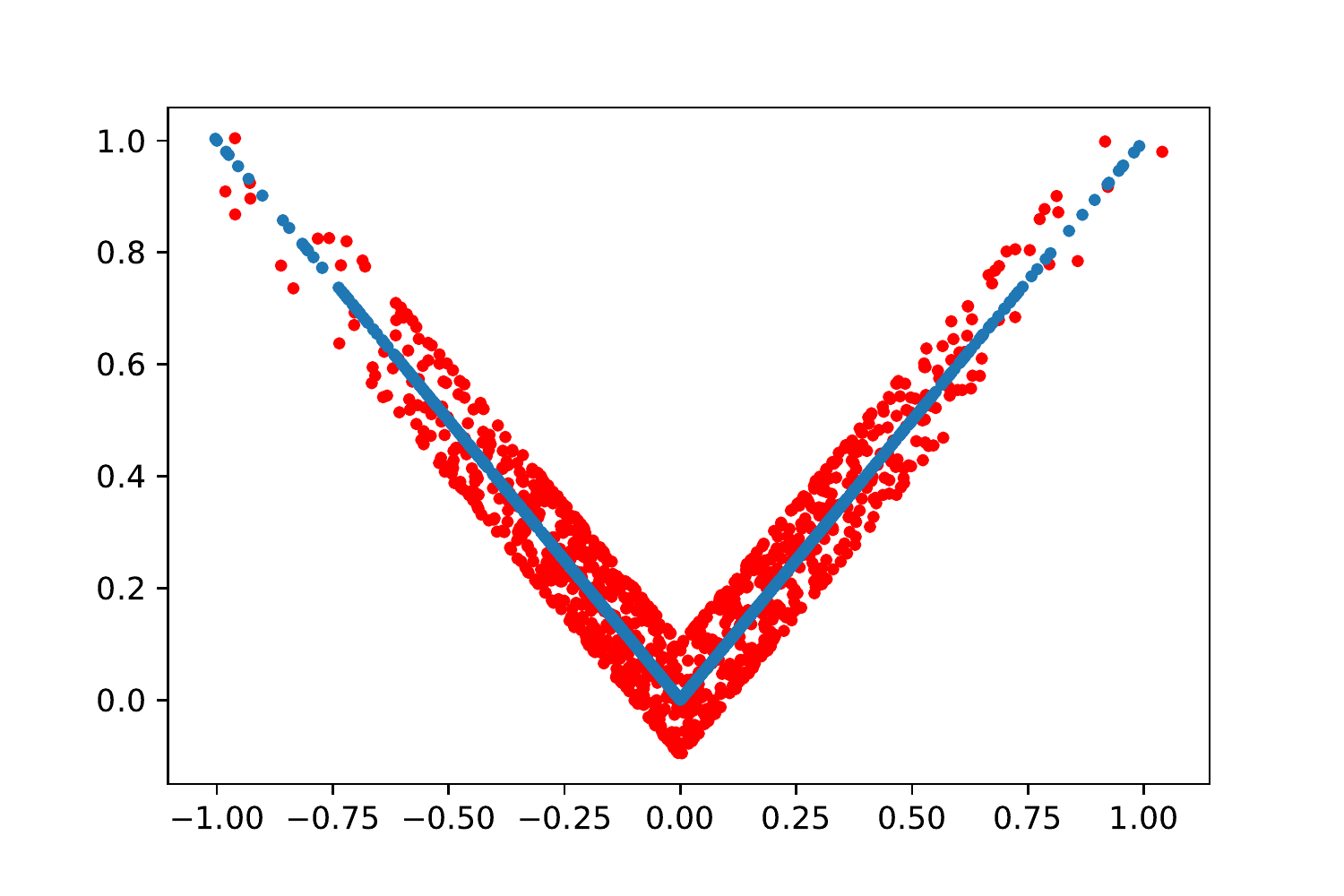}\\
		$x_0$ vs $f(x)$ & $\phi(x)$ vs $F(x)$
	\end{tabular}
\end{figure}

\subsection{MARTHE} The MARTHE dataset \citep{simulationlib} consists of $300$ input-output pairs where outputs are realizations of the numerical simulation of  Strontium-90 transport in the upper aquifer of the RRC ``Kurchatov Institute" radwaste disposal site in Moscow, Russia \citep{volkova2008global}. The input vector has $20$ components describing physical properties of the aquifer, such as hydraulic conductivity, porosity and transversal dispersivity. The outputs are  contaminant concentrations at 10 wells, 
To illustrate our methodology, we use the output at one well, well number eight. 

\begin{figure}[H]
	\centering
	\includegraphics[width=\linewidth]{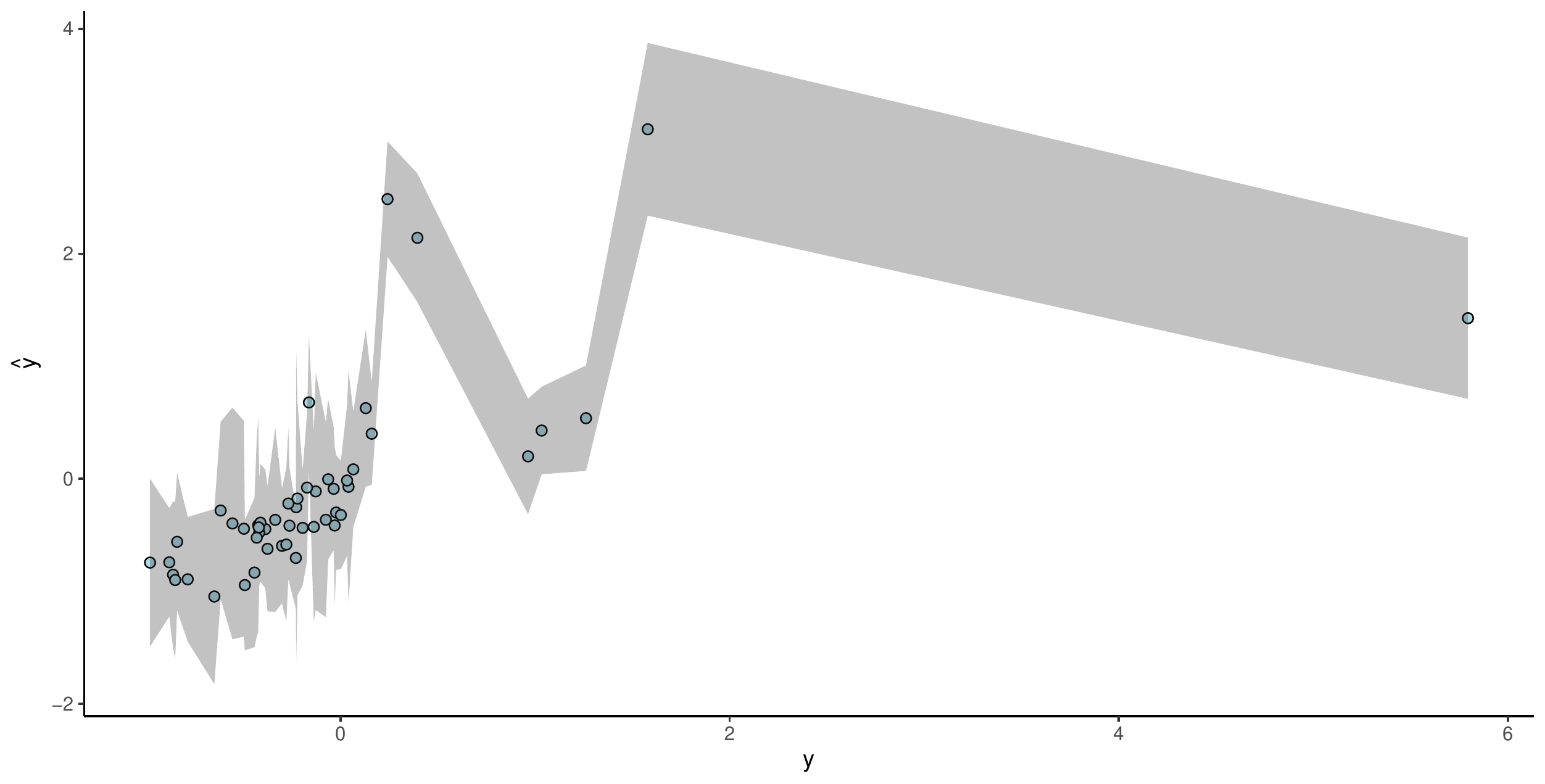}
	\caption{Out-of-sample predictions from plain GP model.}
	\label{fig:mathe-gp}
\end{figure}

Gaussian Processes (GPs) have been used to provide an output map together with uncertainty bands. Figure \ref{fig:mathe-gp} shows the scatter plot of the  predicted values $\hat{Y}$ with 95\% confidence band of the GP model and the actual observations $Y$.

We use a Gaussian Process with zero mean and separable anisotropic Gaussian covariance function which is the sum of inverse exponentiated squared difference plus the nugget
\[
C_{d}(x,x') = \exp\left\{ \sum_{i=1}^p  (x_i - x_i^{'})^2/d_i \right\} + g\delta_{x,x'}
\]
The lengthscale parameters of the covariance function $d_1,\ldots,d_p$ and the nugget parameter $g$ were estimated by maximizing the Bayesian integrated log likelihood \cite{gramacy2016lagp,gramacy2020surrogates}. 

Our merging of two cultures leads to an improved fit and predictive map as follows.  First, we use partial least squares (PLS) \citep{polson2021deep} to provide feature selection.
Remember that the learned features depend on  both the  input and output pairs leading to an optimal mean squared error in-sample fit. Moreover, PLS provides a dimension reduction, that helps in the GP model at the top level. Dimensionality reduction is crucial in many applications. For example in the problems of sequential design of experiment, when lower dimensionality allows for more efficient exportation of the input space \citep{mackay1992information, fedorov1997model}.

We can view our model as a hierarchical latent feature model, where we use the stochastic GP model to provide a predictive distribution at the top level.

\subsection{PLS-GP}
Given a new predictor matrix $X_{*}$ of size $N_*\times p$ we wish to find a predictor of the output $ Y_\star $. 
First, we use the same projection $P$ produces the corresponding input scores $T_{*} = [T_{1,*}, ..., T_{L, *}]$. These can be interpreted as the hidden latent features. 
Then Gaussian process regression finds the predictor $ \hat{U} $ of the output scores $U_{*} = [U_{1,*}, ..., U_{L, *}]$, component-by-component, using the features  $T_{*}  $  as follows
\begin{align*}
\hat Y &= \hat U Q\\
U = & GP(C(t,t'))\\
T &= X P^T
\end{align*}
where $GP(C(t,t'))$'s are the Gaussian process regression predictors. 

\begin{equation*}
\begin{bmatrix}
U\\
U_*
\end{bmatrix}\sim N\left(
\begin{bmatrix}
g\\
g_*
\end{bmatrix},
\begin{bmatrix}
K & K_*\\
K_*^T & K_{**}
\end{bmatrix}\right)
\end{equation*}
where $K = K(t, t)$ is $N\times N$, $K_* = K(t, t_*)$ is $N\times N_*$, and $K_{**} = K(t_*,t_*)$ is $N_*\times N_*$. $K(\cdot,\cdot)$ is a kernel function. 
The conditional mean, $g_*$ from the GP model  is given by
\begin{align*}
g_*(T_*) &= g(T_*) + K_*^TK^{-1}(u - g(T)). 
\end{align*}
Then prediction of $Y$ then takes the form 
$\hat Y_* = \hat U_{*} Q$.

\begin{figure}[H]
\centering
\includegraphics[width=\linewidth]{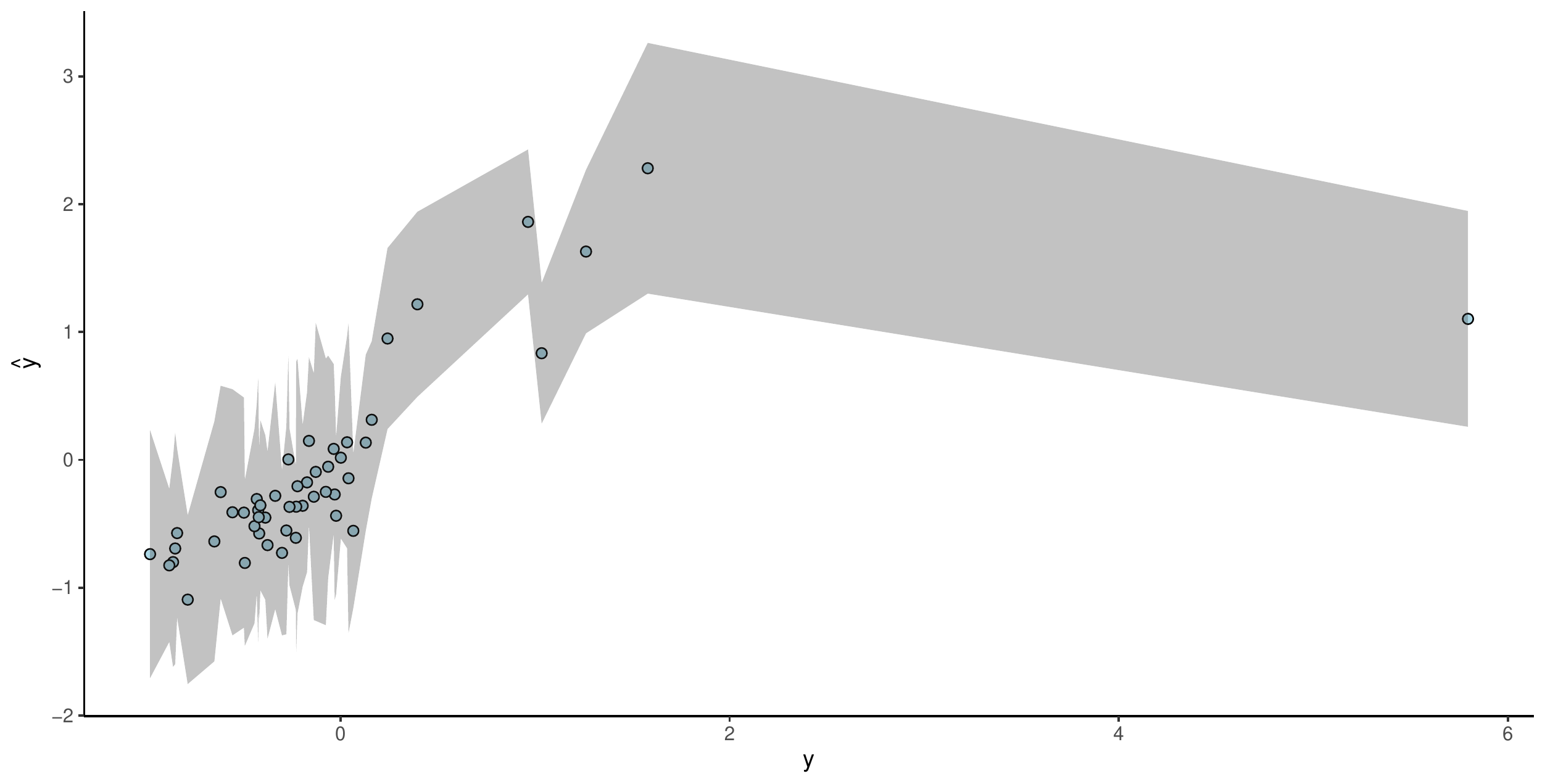}
\caption{Out-of-sample predictions from GP model that uses PLS scores as inputs. Fourteen PLS components were used. }
\label{fig:mathe-gp-pls}
\end{figure}

Figure \ref{fig:mathe-gp-pls} shows the comparison of the predicted $\hat Y$ and actual values $Y$ along with uncertainty bounds

For the PLS dimensionality reduction we used cross-validation to select the number of components.

\subsection{Deep Learning Gaussian Process}
\begin{figure}[H]
	\includegraphics[width=\linewidth]{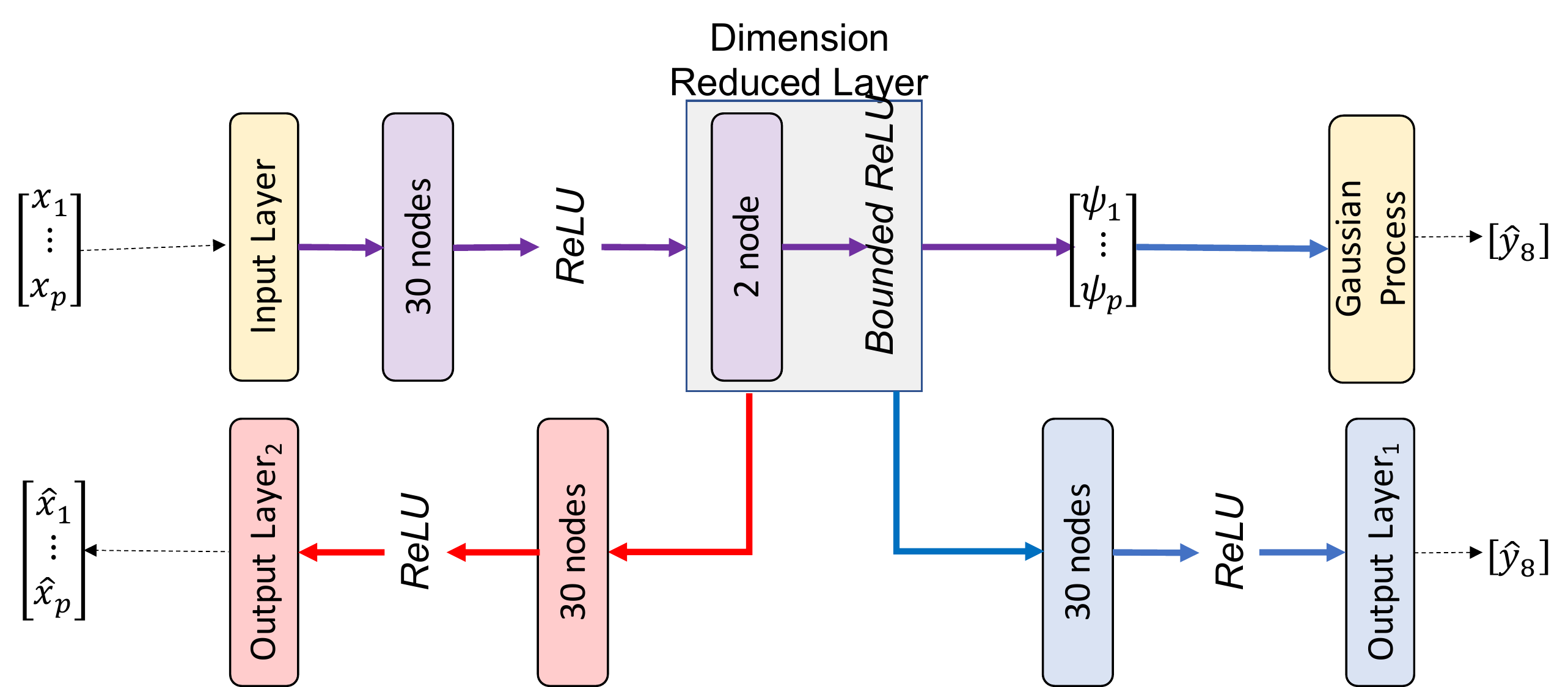}
	\caption{Deep learning model  that consists of the projection function $\phi$ which performs sufficient dimension reduction and predictive rule $F$ that maps lower dimensional representation of inputs to the outputs. }
	\label{fig:dl}
\end{figure}

Both PLS and DL learn a low dimensional representation of the input vector. We build a combined DL-GP model. The goal of the DL model is to find $\psi$, which is a  reduced dimensionality representation of input vector $x$. Then to  use a Gaussian Process to model relations between low dimensional inputs $\psi$ and the output $y$.  Our architecture is shown in Figure \ref{fig:dl}.

\begin{figure}[H]
\centering
\includegraphics[width=\linewidth]{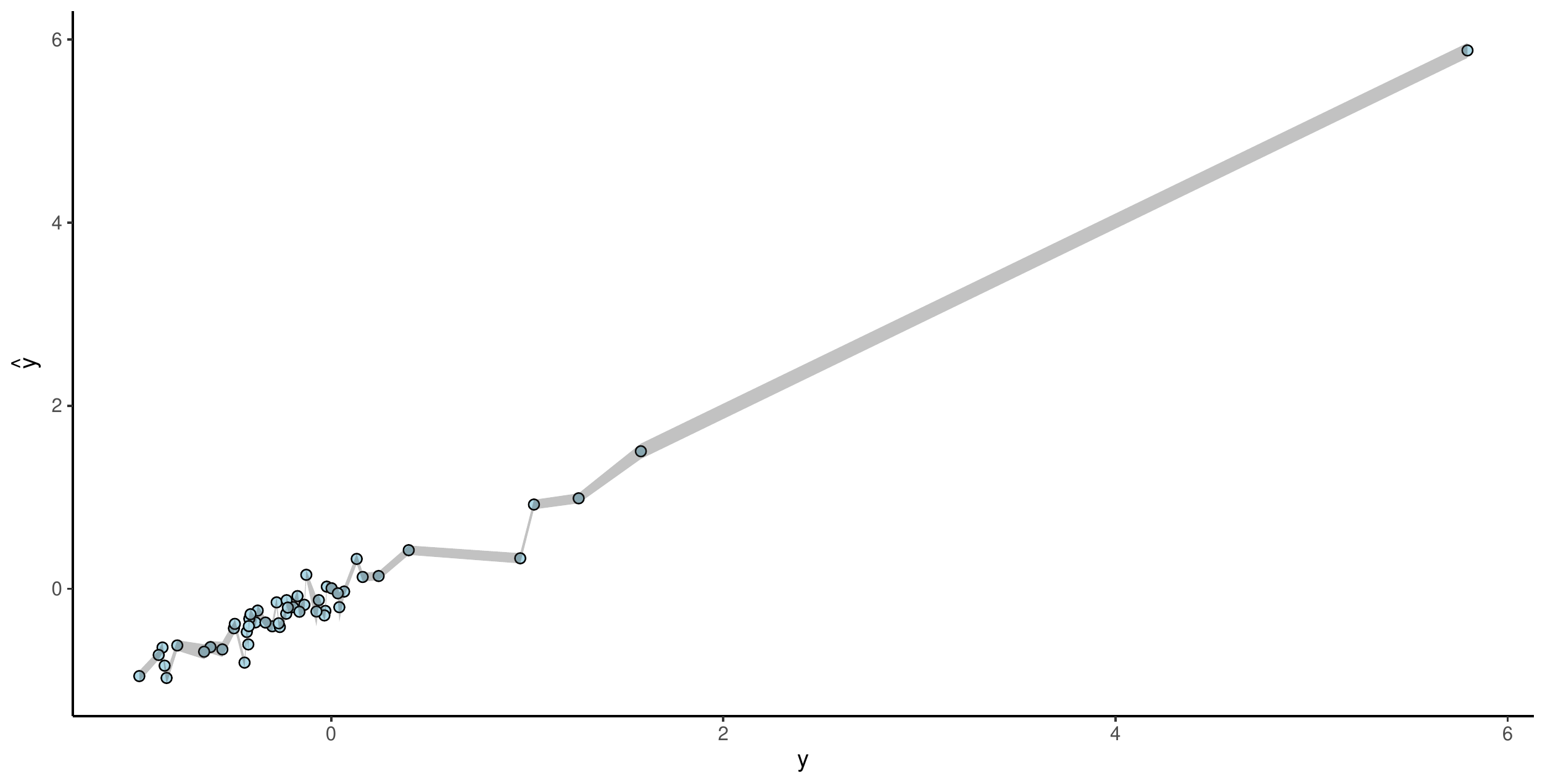}
\caption{Out-of-sample predictions from GP model that uses DL scores as inputs. Ten DL components were used. }
\label{fig:mathe-gp-dl}
\end{figure}

Figure \ref{fig:mathe-gp-dl} shows the predicted and actual scattershot.



\begin{table}[H]
	\centering
\begin{tabular}{l|l|l|l}
& Plain GP & PLS + GP  & DL + GP  \\\hline
RMSE & 4.5      & 1.6      & 0.89    \\
MAPE & 0.8      & 0.73     & 0.16   
\end{tabular}
\caption{Comparison of out-of-sample performance of different models for MARTHE dataset}
\end{table}


\section{Discussion}\label{sec:discussion}
The goal of statistics is to build predictive models along with uncertainty and to develop understanding about the data generating mechanism. Data models are well studied in statistical literature but often do not provide enough flexibility to learn the input-output relations. Black box predictive rules such as trees and neural networks, are more flexible learners but do not provide predictive uncertainties or ability for probabilistic modeling.

Data can be though of as generated by black box on which a vector of input variables $X$  is mapped to an output (or response vector $Y$). One goal is prediction to be able to assign a response variable $Y_*$ to a new (unseen before) input $X_*$. Two cultures have emerged: stochastic methods with parameters or black box predictions rules. What makes a good statistical model? What makes a good prediction rule? Given a model (and computation) leads to an optimal prediction rule. However, in  high dimensional problems finding good models is challenging. One needs a good a priori distribution that gets updated in the light of evidence.

Our methodology provides a merging of these two cultures. We show that deterministic black box rule can be used as transformation of high dimensional inpour and outputs. In the transformed space lead to hidden features that are empirically learned as supposed to theoretically specified. Statistical modeling then provides uncertainty assessment via traditional Bayesian updating methods. 



\bibliography{ref}
\end{document}